\documentclass[twocolumn]{aastex7}

\usepackage{color}
\usepackage{amsmath}
\usepackage{xcolor}
\usepackage{mathrsfs}
\usepackage{natbib}
\usepackage{gensymb}
\usepackage{appendix}
\usepackage[T1]{fontenc}
\usepackage{graphicx}   
\usepackage{float}
\usepackage{adjustbox}
\usepackage{tabularx}
\usepackage{booktabs}

\defcitealias{Kerr21}{SPYGLASS-I}
\defcitealias{Kerr22b}{SPYGLASS-III}
\defcitealias{Kerr22a}{SPYGLASS-II}
\defcitealias{Kerr23}{SPYGLASS-IV}
\defcitealias{spyglass_5}{SPYGLASS-V}
\defcitealias{Kerr25}{SPYGLASS-VI}
\defcitealias{Kerr25b}{SPYGLASS-VII-A}

\begin{document}

\title{From Dense Cores to Young Stars: The Ages, Velocities, and Dynamics of $\rho$ Ophiuchi}

\author[0009-0002-2836-3447]{Sharone E. Abhilash}
\affiliation{Department of Chemical \& Physical Sciences, University of Toronto Mississauga\\ 3359 Mississauga Road, Mississauga, ON L5L 1C6, Canada}
\affiliation{Dunlap Institute for Astronomy \& Astrophysics, University of Toronto\\
Toronto, ON M5S 3H4, Canada\\}
\email{sharone.abhilash@mail.utoronto.ca}

\author[0000-0002-6549-9792]{Ronan Kerr}
\affiliation{Dunlap Institute for Astronomy \& Astrophysics, University of Toronto\\
Toronto, ON M5S 3H4, Canada\\}
\email{ronan.kerr@utoronto.ca}

\author[0000-0002-1379-4204]{Alexandre Roman Lopes}
\affiliation{Departamento de Astronom\'ia, Universidad de La Serena \\ Av. Ra\'ul Bitr\'an 1305, La Serena 1700000, Coquimbo, Chile}
\email{aroman@userena.cl}

\author[0000-0002-2269-9348]{Sergio S\'anchez-Sanju\'an}
\affiliation{Instituto de Astronom\'ia, Universidad Nacional Aut\'onoma de M\'exico \\ Apartado Postal 106, C. P. 22800, Ensenada, B. C., M\'exico}
\email{sergsanchez@astro.unam.mx}

\author[0000-0001-7998-226X]{Lisa Prato}
\affiliation{Steward Observatory, University of Arizona, 933 North Cherry Avenue, Tucson, AZ 85721, USA}
\affiliation{Space Science Institute, 1601 29th St. Suite 1292, Boulder, CO 80301, USA}
\email{science.prato@gmail.com}

\author[0000-0001-8164-653X]{Travis ~A. Rector}
\affil{Department of Physics and Astronomy, University of Alaska Anchorage, Anchorage, AK 99508, USA}
\email{tarector@alaska.edu} 

\author[0000-0003-3526-5052]{Jos\'e G. Fern\'andez-Trincado}
\affiliation{Centro de Investigaci\'on en Astronom\'ia, Facultad de Ingenier\'ia, Ciencia y Tecnolog\'ia, Universidad Bernardo O'Higgins \\ Av. Viel 1497, Santiago 8370993, Chile}
\email{jose.fernandez@ucn.cl}

\author[0000-0003-1805-0316]{Richard R. Lane}
\affiliation{Centro de Investigaci\'on en Astronom\'ia, Facultad de Ingenier\'ia, Ciencia y Tecnolog\'ia, Universidad Bernardo O'Higgins \\ Av. Viel 1497, Santiago 8370993, Chile}
\email{richard.lane@ubo.cl}

\author[0000-0002-3481-9052]{Keivan G. Stassun}
\affiliation{Department of Physics and Astronomy, Vanderbilt University, Nashville, TN 37235, USA\\}
\email{keivan.stassun@vanderbilt.edu}

\author[0000-0002-5365-1267]{Marina Kounkel}
\affiliation{Department of Physics and Astronomy, University of North Florida, 1 UNF Dr, Jacksonville, FL, 32224, USA}
\email{marina.kounkel@unf.edu}

\begin{abstract}
The $\rho$ Ophiuchi star forming complex ($\rho$ Oph) is one of the nearest and most active sites of ongoing star formation, comprising multiple clouds and substructures differentially exposed to stellar feedback, making it a prime case study for feedback-driven compression, sequential star formation, and cloud-cloud interactions shaping diverse stellar populations. We present a demographic and kinematic study of $\rho$ Oph, combining catalogs of protostars, dense cores, and pre-main-sequence stars into a unified catalog spanning 17 subregions, and some of the regions, like L1689 and L1688, are decomposed into on- and off-cloud components. Using protostellar class lifetimes and isochronal fitting, we derive ages across the complex, revealing a west-to-east gradient spanning ~1-8 Myr, with off-cloud populations typically older by $\sim1$ Myr. Using transverse velocities and dynamical traceback, we find that compression arcs within L1688 identified by previous work are kinematically distinct and may be remnants of separate star formation episodes.  Similar ages but distinct kinematics within the arcs suggest an ongoing collision, potentially contributing to more star formation and the large stellar population of $\rho$ Oph. More broadly, heterogeneous velocity patterns across the complex suggest that feedback, cloud-cloud interactions and local cloud structure collectively shape its dynamical evolution. These results suggest a single, feedback-driven sequence is insufficient for the dynamical complexity of environments like $\rho$ Oph. 



\end{abstract}

\section{Introduction}

Stars form in groups within giant molecular clouds. Following the dispersal of their natal clouds, these groups evolve into stellar associations: gravitationally unbound ensembles of stars that remain physically and kinematically related \citep{lada_lada2003}.

One of the most well-studied, low-mass star-forming regions in the solar neighborhood is the $\rho$ Ophiuchi complex, located $\sim 139$~pc away \citep{Zucker_2020}, with a total gas mass estimated at $7.4\times10^{3}\,M_{\odot}$ \citep{Ridge_2006}.  It is located within the Upper Scorpius (USco) subgroup, the youngest component of the larger Scorpius–Centaurus (Sco-Cen) stellar association \citep{Armstrong_2025}. The Ophiuchus dust cloud is described by \citet{Loren_1989} as a network of ‘cobwebs’ centered on two main clouds: L1688 and L1689. Of these, L1688 hosts the bulk of active star formation, in contrast to the more quiescent L1689 cloud dubbed “the dog that didn’t bark” \citep{Nutter_2006} which lies to its east. Several filamentary extensions, including L1709, L1740, L1744, L1755, L1765, and L1712 also extend from L1688 and host smaller populations of young stellar objects (YSOs) \citep{Nutter_2006}.  

Due to its proximity, $\rho$ Oph has been the prime target for extensive multi-wavelength observations that have uncovered a wide range of evolutionary stages, including dense cores \citep[e.g.,][]{Ladjelate_2020}, embedded Class~0/I protostars and more evolved Class~II and III objects \citep[e.g.,][]{Dunham_2015, Pokhrel_2023, Droxo_2010}. Rector et al. (in prep.) also recently detected numerous new Herbig–Haro objects in the region. Various complementary gas maps (e.g., the COMPLETE survey; \citealt{Ridge_2006}, and the Green Bank Ammonia Survey; \citealt{Friesen_2017}) also trace cloud spatial and kinematic structure of $\rho$ Oph. 

Associations like Sco-Cen frequently exhibit internal age gradients (e.g., Rahaman et al. (in prep.);\citealt{Kerr_2024}; \citealt{Zari_2018},) reflecting spatially extended and dynamically complex formation histories \citep{Armstrong_2025}. Because these systems are young and only partially dynamically evolved, they retain the spatial and kinematic signatures of their formation \citep{lada_lada2003}. In this way, stellar associations not only provide a snapshot into star formation but a record of the events that shaped them \citep{Kerr21}.

Observations across various star-forming regions have revealed a surprising diversity in star formation histories. Feedback from massive OB stars in associations, through radiation, winds and supernovae, can disperse gas, but also generate shocks and compress neighboring regions of the cloud. Sequential star formation instigates subsequent waves of collapse that cascades throughout the larger cloud leading to successive generations of star formation \citep{Elmegreen_1977}. Some clouds evolve through multiple, overlapping episodes of star formation (e.g., Taurus; \citealt{Kerr21}), while others show signs of cloud-cloud collisions (e.g., Circinus; \citealt{Kerr_Cir_2025}). Disentangling these different modes requires combining spatial, kinematic and temporal analysis to reconstruct a region's evolutionary history since no single diagnostic uniquely determines the underlying formation history.

Within Sco-Cen, several studies have identified observational signatures of sequential star formation. \citet{Posch_2025} show that the USco subgroup hosts a series of stellar populations with increasing bulk velocities, decreasing ages and masses, signatures consistent with a feedback-driven sequential star formation. At the forefront of this chain is $\rho$ Oph, long suspected to have formed via feedback-triggered star formation influenced by massive stars such as $\sigma$ Sco, $\alpha$ Sco, 22 Sco and the $\rho$~Oph star \citep{Loren_1986, Motte_1998}, with compression fronts consistent with feedback-driven origin \citep{Ladjelate_2020}. \citet{Posch_2025} suggest that $\rho$~Oph represents not only the youngest product of the existing sequence but also the first link in a new wave of star formation propagating through the region. Its combination of YSO density, heterogeneity, proximity, and rich multiwavelength coverage makes $\rho$ Oph an archetype for studying the earliest chapters of star formation and reconstructing its temporal, spatial, and kinematic evolution.

In this study, we reconstruct the spatial and temporal star formation history of the $\rho$ Oph cloud complex by integrating a broad entourage of stellar, protostellar, molecular gas, radial velocity datasets described in \ref{section: data}. We begin by outlining our cross-matching procedure and the construction of the master catalog in Section \ref{methods}. In Section \ref{subsection:ages}, we derive  age estimates by combining both isochronal and population-based metrics (e.g., protostellar lifetimes and abundances), while our kinematic analysis of stellar subgroups is presented in Section \ref{sec: on and off cloud}. Finally, we discuss the implications of our results, the relation of substructure to proposed feedback sources in Section \ref{dis} and summarize our conclusions in Section \ref{sec:conc}.

\section{Data}
\label{section: data}

\subsection{The SPYGLASS Catalog}
\label{subsection:spyglass}
The SPYGLASS program provides one of the most comprehensive censuses of photometrically young stars within 1 kpc of the Sun by leveraging {\it Gaia} astrometry and photometry in combination with multi-survey spectroscopic youth indicators \citetalias{Kerr23}. Using a Bayesian framework to evaluate stellar youth probabilities, SPYGLASS identifies a large population of photometrically young stars and organizes them into coherent kinematic structures. The SPYGLASS catalog provides stellar parameters such as photometric youth probabilities, association membership probabilities, and {\it Gaia} DR3 astrometry (positions, proper motions, and distances from \citet{Bailer_2021}) and photometry ($G$, $G_{\mathrm{BP}}$, $G_{\mathrm{RP}}$). This makes SPYGLASS particularly well suited for identifying pre-main sequence populations and characterizing their kinematics. 

\begin{figure*}[t]
    \centering
    \includegraphics[width=\textwidth]{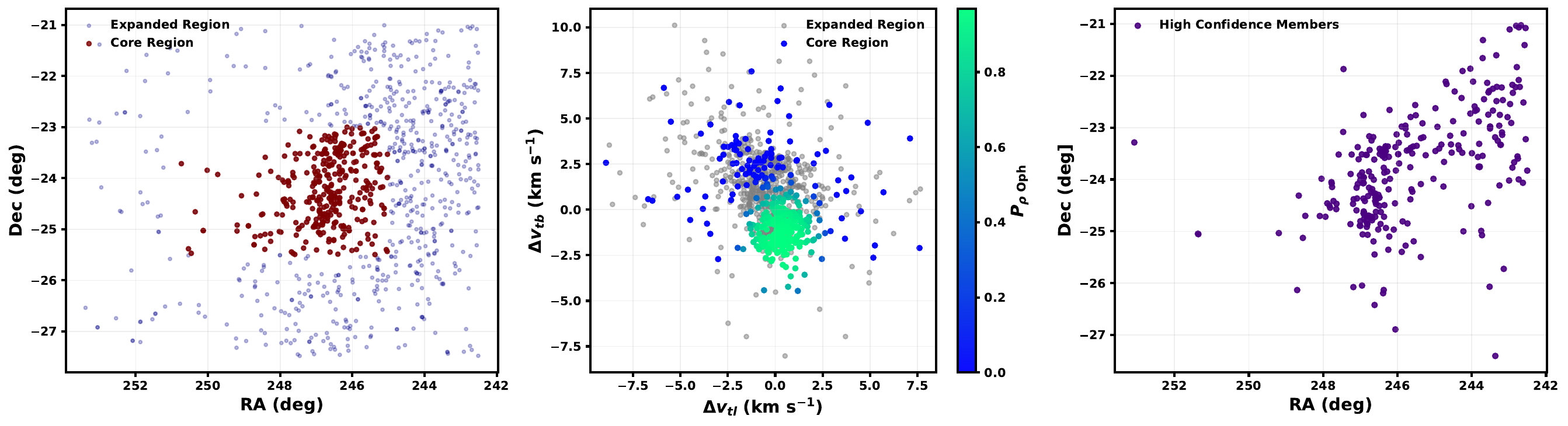}
    \caption{The leftmost panel shows the core $\rho$ Oph region (RA = 245.0-251.0°, Dec = -25.5° to -23.0°), and the extended region (RA = 242.5-253.5°, Dec = -27.5° to -21.0°) defined by the spatial extent of the 13CO emission from the COMPLETE survey. In the middle panel, we show the distribution of the sources in transverse anomaly space, where the expanded sample is shown in gray and sources within the core region are coloured by their GMM membership probability. The rightmost panel shows the final spatial distribution of selected members after applying all kinematic and quality cuts, including the transverse velocity selection, RV constraints, spatial-photometric membership probability.}
    \label{fig:spyglass}
\end{figure*}

\citetalias{Kerr21} identified numerous YSOs associated with the $\rho$ Oph population as part of the broader Sco-Cen association. Subsequent work in \citetalias{Kerr23} expanded the extent of Sco-Cen, connecting its substructures to the adjacent Chamaeleon complex and other populations. The Sco-Cen complex is given the catalog ID SCYA-103 and it consists of 8157 photometrically young members. To isolate $\rho$ Oph from the much larger Sco-Cen association and assess the velocity distribution, we define an inner, compact region identified in \citetalias{Kerr21}, and the extended region defined by the bounding box that follows the extent of of the 13CO emission maps from the COMPLETE Survey \citep{Ridge_2003}. We highlight the two regimes in Figure \ref{fig:spyglass} where the compact region serves as a sample of high probability young members in $\rho$ Oph, while the extended region allows us to search for more spatially distributed stars sharing the same kinematic features. 

To search for kinematically linked members of $\rho$ Oph, we use the transverse velocity anomaly, defined as the difference between each star's velocity and that expected from the median Sco-Cen UVW vector \citep{Kerr23}. This metric is useful because it removes the position-dependent apparent motion that a common 3D bulk velocity We model the distribution of transverse velocity anomalies in the compact $\rho$ Oph core using a two component Gaussian Mixture Model, implemented with \texttt{sklearn.mixture.GaussianMixture} (GMM) \citep{scikit-learn}. The results are shown in the middle panel of Figure \ref{fig:spyglass}, where a clear bimodal structure is evident within the velocity anomaly space describing the core region. This bimodality motivates our adoption of a two-component GMM. We identify the $\rho$~Oph component as the more compact of the two, i.e. the one with the smaller total velocity dispersion $\sqrt{\sigma_{tl}^2 + \sigma_{tb}^2}$. We then retain stars from the core region with a posterior probability of $P>0.5$ of belonging to the $\rho$ Oph population and serve as our reference population. The bivariate Gaussian fit by the GMM was applied to the extended region to evaluate membership probabilities. Stars lying within the $2\sigma$ contour of this distribution are classified as high-confidence $\rho$ Oph members.

Radial velocities (RVs) are imperative for defining membership and for thorough kinematic analysis and therefore we also refine our sample based on RVs. We remove extreme outliers by restricting our analysis to stars with RVs in the range of $-60$ to $+60$ km s$^{-1}$ and constructed the distribution using a bin width determined by the Freedman-Diaconis rule. We modeled the central peak of the distribution with a Gaussian profile, while a background level was estimated from the outer bins. We retain stars within $3\sigma$ of the fitted mean. We exclude stars whose RV uncertainties exceed $3\sigma_{v_t}$, where $\sigma_{v_t}$ is the dispersion of the total transverse velocity of the $\rho$~Oph component, although the corresponding stars are not removed from the catalog. 

Our final assessment of membership is the re-normalized spatial-photometric membership probability $P_{\mathrm{sp}}$, which reweighs the photometric youth probability by accounting for the local overabundance of young stars relative to that expected from a uniform star formation history \citep{Binney_2000}. This metric is primarily intended to suppress contamination from background and foreground stars rather than to uniquely identify $\rho$ Oph members. 

We retain only stars with $P_{\mathrm{sp}} > 0.5$. The 260 stars that survived our spatial and kinematic restrictions define our final high-confidence $\rho$ Oph members which we highlight in Figure \ref{fig:spyglass}.

\subsection{Protostars \& DiskDominated YSOs}
\label{subsection: protostars}

Protostars represent actively forming stars in the accretion phase, consisting of a hydrostatic object still embedded within and accreting material from an infalling envelope \citep{Andre_2000}. Their evolutionary stages are classified by observational features \citep{Heimsoth_2022} where Class 0 protostars represent the earliest stage, undergoing rapid accretion from a dense, massive envelope lasting approximately 0.15-0.24 Myr, while in class I sources, accretion slows as it gathers material from remaining envelope lasting 0.31-0.48 Myr \citep{Dunham_2015}. As the system evolved into the more optically visible Class II phase, the envelope has largely dispersed, leaving behind an accretion disk, with a  median disk lifetime of 2–3 Myr \citep{Williams_2011}. In the subsequent Class III stage, the system transitions to a weak-lined T-Tauri phase, where accretion has largely ceased and the disk has dissipated or is in the final stages of dispersal \citep{Feig_2010, Williams_2011}. Distinguishing between these different stages is important as their relative numbers and spatial distribution provide clues in resolving temporal star formation histories. In the following subsections we describe the three protostellar catalogs (eHOPS, SESNA and Dunham) included in this work. 

\subsubsection{eHOPS}
\label{subsubsection: eHOPS}
The extended Herschel Orion Protostar Survey (eHOPS) provides a comprehensive multiwavelength catalog designed to identify and characterize deeply embedded YSOs. eHOPS assembles 1-850 $\mu$m spectral energy distributions (SEDs) by combining near-IR 2MASS photometry \citep{2MASS_2006}, mid-IR Spitzer/IRAC \citep{IRAC_2004} and MIPS \citet{MIPS_2004} photometry from the SESNA archive, Spitzer/IRS spectra, far-IR Herschel/PACS and SPIRE photometry, and submillimeter JCMT/SCUBA-2 fluxes. A detection in at least one Herschel/PACS band (70, 100, 160 $\mu$m) is required. Because of this far-IR coverage, eHOPS is especially important in identifying the deeply embedded Class 0 and I sources. 

The classification scheme integrates multiple characteristics: two point mid-IR SED(between 4.5 $\mu$m and 24 $\mu$m) based on \citet{Greene_1994} and \citet{Evans_2009}, Spitzer color cuts, far-IR flux ratios, and completeness limits for bands with non detections. These combined criteria allow eHOPS to separate protostellar candidates (Class 0, Class I and Flat-spectrum sources) from more evolved pre-main-sequence stars. For each source, the catalog provides fluxes, IRS spectra when available, derived SED properties including bolometric temperature ($T_{\mathrm{bol}}$), bolometric luminosity ($L_{\mathrm{bol}}$), extinction ($A_V$) and envelope mass, and an evolutionary class: Class 0, Class I or Flat-spectrum for protostellar candidates, and ``Disk'' or ``Reddened Disk'' for more evolved sources. These quantities are particularly useful for resolving classification ambiguities among the youngest YSOs: $T_{\mathrm{bol}}$ and the evolutionary class underpin the protostellar age assignments described in Section~\ref{subsection:ages}.

\subsubsection{SESNA}
\label{subsubsection: SESNA}

The Spitzer Extended Solar Neighborhood Archive (SESNA; R.A.Gutermuth et al., in prep.) provides uniform photometry from 1- 24 $\mu$m for nearby star forming regions like Orion B, Lupus, Mon R2, Perseus (across > $90~\mathrm{deg}^2$), combining near-infrared data from 2MASS \citep{2MASS_2006}, with Spitzer IRAC \citep{IRAC_2004} and MIPS \citep{MIPS_2004}. SESNA has uniform observing parameters and applies a uniform photometric extraction and reduction procedure across all surveyed regions, improving consistency relative to earlier Spitzer-based YSO compilations \citep{Pokhrel_2023}. SESNA provides positions, fluxes, uncertainties, and evolutionary classes for the YSOs. 

\subsubsection{The Dunham Catalog}
\label{subsection: dunham}

The catalog of \citet{Dunham_2015} presents a uniform list of YSOs identified in 18 molecular clouds observed by the Spitzer "Cores to Disk" (c2d) and 'Gould Belt' surveys. For each source, \citet{Dunham_2015} assembled SEDs from 2MASS + Spitzer/IRAC/MIPS photometry and correct them for foreground extinction. The Dunham catalog provides the extinction-corrected spectral index computed via a least squares fit to all available 2-24$\mu$m. The different sources were then classified based on their extinction corrected spectral index as outlined by \citet{Greene_1994} and \citet{Evans_2009}. They also provide flags for highly likely background AGB stars and whether or not each YSO is associated with a dense core as traced by sub-millimeter or millimeter continuum emission.  The three catalogs differ in wavelength coverage, extinction treatment and classification diagnostics. These differences are expected, and we describe how we reconcile them into a single evolutionary sequence in Section \ref{subsection: cross_matching}.

\subsection{The Prestellar and Protostellar Dense Core Catalogs}
\label{subsection: densecores}
Stars form within dense cores. These cores represent compact, gravitationally bound condensations thought to be the direct progenitors of protostars \citep{Andre_2001}. To trace these earlier stages in $\rho$ Oph, we incorporate a catalog of dense cores derived from the Herschel Gould Belt Survey (HGBS; \citealt{Andre_2010}; \citealt{Ladjelate_2020}). Observed with Herschel's PACS and SPIRE instruments, dense cores can locate cold, high optical depth environments where stars from. The HGBS observed the L1688 and L1689 regions of Ophiuchus, and the catalog presented by \citet{Ladjelate_2020} includes core mass, temperature, radius (in pc), spatial positions and core type. We also include the JCMT Gould Belt Survey which presents SCUBA-2 observations of the Ophiuchus molecular cloud \citep{Pattle_2015}. This survey identifies these sub millimeter sources as either protostellar or starless cores. It combines SCUBA-2 data with molecular line observations from HARPS, as well as ancillary Herschel and IRAM data. Important for this work, the JCMT GBS provides a coverage of L1709 and L1712 regions of Ophiuchus. In this work, we include robust and candidate prestellar cores, which are gravitationally bound (or tentatively bound) and expected to collapse and form stars, and protostellar cores, which already contain a YSO. We exclude unbound starless cores, which are not expected to form stars. Including starless unbound cores may introduce a population that is unrelated to the ongoing or future star formation of the region. The catalog therfore contains a total of 193 dense cores: 116 prestellar tentative bound cores, 28 robust prestellar cores, and 49 protostellar cores. We added 2 additional sources from \citet{Pattle_2015}; 2 protostellar cores (L1709 and L1712).

\subsection{The COMPLETE 13CO Molecular Gas Survey}
\label{subsection: gas}

The material required to form stars in clusters is drawn from a reservoir of molecular gas, and the resulting star formation rate is therefore closely tied to physical properties of the parental molecular cloud. In this work, we make use of 13CO (U = 1 -> 0) molecular line data from the COMPLETE Survey \citep{Ridge_2006}. The data provide a position-position-velocity (PPV) cube (RA, Dec, $v_{\rm LSR}$) of the gas, where $v_{\rm LSR}$ is the line-of-sight velocity with respect to the Local Standard of Rest, obtained using the 14-m Five College Radio Astronomy Observatory. The $^{13}$CO transition has a lower optical depth than $^{12}$CO, allowing it to probe deeper into molecular clouds. While it can become optically thick in the densest cluster centers, it remains an effective tracer of the broader cloud structure, particularly in moderately dense and more extended regions \citep{Ridge_2003, Narayanan_2008}. The gas kinematics are particularly valuable for studying embedded populations, where stellar radial velocities are often unavailable; in such cases, $^{13}$CO velocities can serve as useful proxies for stellar RVs. For each core, we adopt the centroid velocity of a Gaussian fit to the $^{13}$CO spectrum at the core position. These velocities are provided in the LSRK frame and are converted to the heliocentric frame. 

\subsection{Kinematic Data}
\label{subsection: kinematic_data}
The Montreal Open Clusters and Associations database (MOCAdb) \citet{moca} is a compilation of data for stellar associations within 500 pc, which unifies data from Gaia, Hipparcos, 2MASS, and wide range of found based surveys. It includes measurements for parallax, proper motion, and RVs, from literature identified $\rho$ Oph members. For this work, we query the MOCAdb entry 'ROPH' to gather positions, proper motions (including IR derived measurements), and RVs, \textit{Gaia} DR3 IDs, along with uncertainties, and BANYAN associations and probabilities. 

We also include an updated form of the RV database introduced in \citet{spyglass_5}. This dataset contains measurements from a broad assortment of RV surveys covering the solar neighborhood, most notably GALAH DR4 \citet{Buder_2025}, APOGEE \citet{APOGEEDR17}, and the Gaia-ESO Survey \citep{Randich_2022}, in addition to several smaller surveys. We also include two additional sources to supplement the literature sources. First, we introduce new RVs from SDSS-V \citep{Kollmeier26} taken with the APOGEE spectrograph \citep{APOGEE} at the Las Campanas Observatory's 2.5m du Pont Telescope \citep{DuPontTelescope}. APOGEE provides near-IR coverage with $R\sim22,500$, making it well-suited for high-quality RVs. The $\rho$ Oph Complex has been covered extensively as part of SDSS-V's Milky Way Mapper (MWM) survey, which observes young stellar objects across several nearby star-forming regions \citep{Kounkel23}. Some APOGEE RVs we include here come from spectra published in SDSS DR19 \citep{SDSSDR19}, but we also include unpublished spectra that we have access to through the SDSS-V collaboration. 

Our second new dataset comes from the Anglo-Australian Telescope (AAT), taken through NSF NOIRLab proposal 2021A-0265, which used the HERMES spectrograph and Two Degree Field System (2dF) \citep{AAT2DF, HERMES} to observe candidate young stars in several fields spread across Sco-Cen. These observations were reduced and published alongside GALAH DR4 \citep{Buder_2025}, but this is the first paper where results from this program have been independently analyzed.

\section{Methods}
\label{methods}
\subsection{Protostar Cross-matching}
\label{subsection: cross_matching}

The eHOPS, SESNA, Dunham and SPYGLASS catalogs each identify YSOs in $\rho$ Oph but differ in wavelength coverage, classification schemes, and YSOs of interest. To build a merged catalog, we establish a uniform framework that resolves duplicates and conflicting class assignments within the catalogs. We adopt a hierarchy based on the diagnostic information each catalog provides. Where available, we use the eHOPS classification, which gains substantial leverage from far-IR coverage of the envelope emission; otherwise we use the extinction-corrected spectral indices of \citet{Dunham_2015}; and for the remaining sources we use the homogeneous mid-IR classification from SESNA.

Because eHOPS and some Dunham sources include far-IR and extinction-corrected quantities that are unavailable in SESNA, we adopt their Class 0, I and Flat-spectrum designations whenever present. Roughly half of the classification disagreements between the three catalogs involve Flat-spectrum sources ($-0.3 < \alpha_{4.5-24} < 0.3$), which lie between the Class I and Class II populations. This SED class can arise from several physical situations, including protostars viewed at favorable inclinations, sources with tenuous residual envelopes and heavily reddened disks \citep{Pokhrel_2023}, and does not map uniquely onto a single evolutionary stage \citep{Dunham_2015, Heiderman_2015}. Flat-spectrum sources nonetheless show accretion signatures distinct from typical Class II disks \citep{Doppmann_2005}, so we retain them as a separate intermediate class, adopting the Flat-spectrum classification where SESNA assigns Class I or Class II. We test the sensitivity of our ages to this choice in Section \ref{subsection:ages}.

For more evolved YSOs, SESNA and Dunham offer finer distinctions than the broader stamp of 'Disk' and 'Reddened Disk' categories used in eHOPS. When SESNA and Dunham agree on a Class II or III classification, we adopt their assigned class and retain the SESNA source. When SESNA and Dunham disagree, we adopt the Dunham classification. We also reclassify four eHOPS 'Disk' sources using a two point $4.5$–$24\mu$m spectral index (3 Class IIs and 1 Class III). 

Next, we cross match the combined eHOPS and SESNA catalog with the Dunham catalog after removing sources flagged as AGB stars. The Dunham catalog had 277 sources (174 Class IIs, 43 Flat, 34 Class III, 26 Class 0 + I). Class 0 + I is differentiated using the extinction corrected bolometric temperature. This second cross-match resulted in 257 matches and 20 unmatched Dunham sources. Finally, the merged catalog catalog is cross matched to 260 SPYGLASS sources, resulting in 48 matches. For these, we retain the SPYGLASS source and the astrometry that accompanies it but preserve the IR based evolutionary class designation. The remaining unmatched SPYGLASS sources are labeled as 'PMS' to denote a broader pre-main sequence population. 

Following our catalog matching and class conflict resolving prescription, we merge the catalog using reciprocal nearest neighbor matching within a 5'' radius of the sources. First, we match between 51 eHOPS protostars (6 Class 0, 19 Class I, 25 Flat, and 1 Class II), 124 eHOPS disk-dominated sources (95 Disk, 29 Reddened Disk) and 386 SESNA sources (271 Class II, 82 Class I, 23 Class III, and 10 Class 0 sources). This cross match resulted in 137 mutual identifications leaving 38 unmatched eHOPS sources and 249 unmatched SESNA sources. 

\begin{figure*}[t]
    \centering
\includegraphics[width=0.99\textwidth]{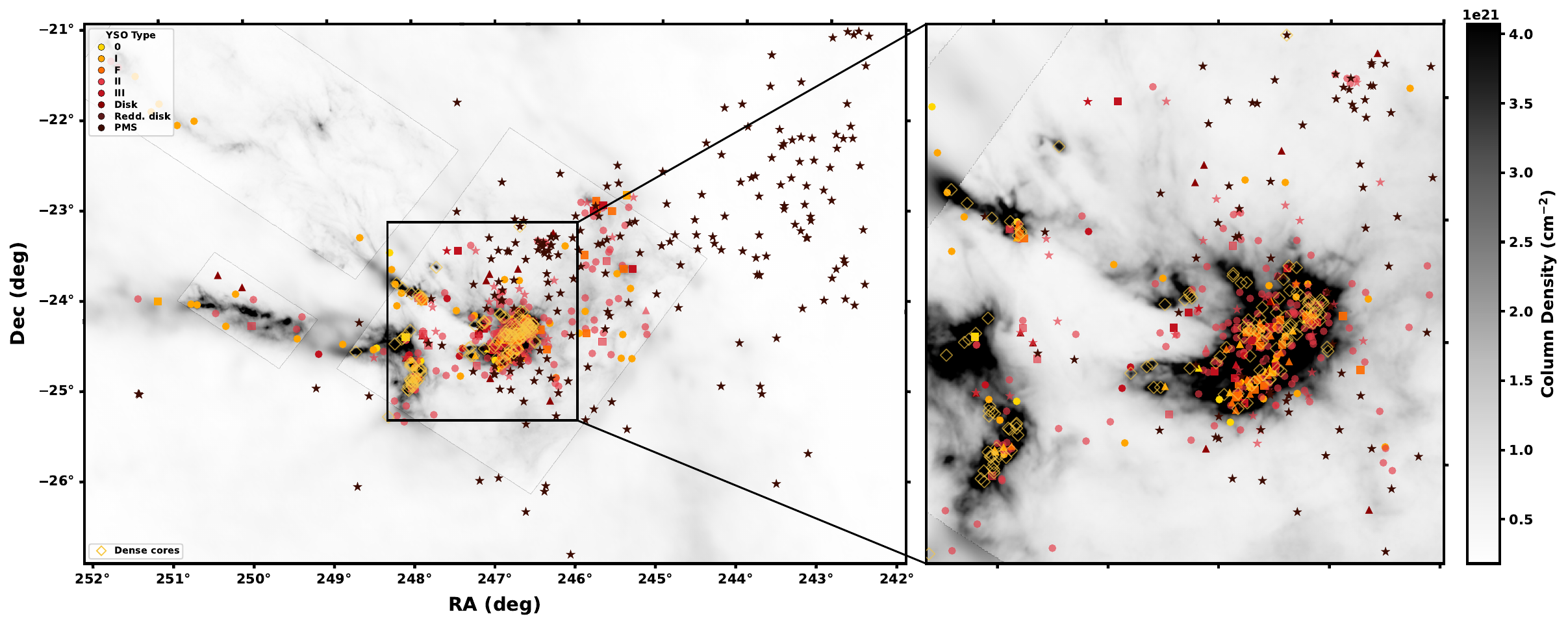}
    \caption{The spatial distribution of sources in $\rho$ Oph, with colour denoting the category of the YSO and marker shapes indicating the source catalog and dense cores. The left panel shows the full extend of the YSOs and the region spanned, while the right panel shows a zoomed-in view of the main L1688 cloud. The background combines a high resolution Herschel column density map of the L1688, L1712, North Streamer fields \citep{HOTT_2022} embedded within a lower-resolution Planck dust map \citep{Planck_2013}.}
    \label{fig:all}
\end{figure*}

The resulting catalog, shown in Figure \ref{fig:all}, contains 212 PMS stars from SPYGLASS, 37 Class III, 250 Class II, 53 Flat, 70 Class I, 16 eHOPS Disks, 4 Reddened Disks, 14 Class 0. The catalog includes 247 sources originating from SESNA, 212 from SPYGLASS, 72 from eHOPS, 69 from Dunham. We also cross match all our protostars with the dense core catalog. We matched each YSO to the nearest core within the core's derived radius. Of the 49 protostellar cores, 48 have an associated protostar. We also find protostar associations for 11 tentative bound prestellar cores and 4 prestellar cores, flag each accordingly in our final catalog. See Table \ref{tab:yso} for reference.

\begin{deluxetable*}{llccccclll}
\tablecaption{Source Catalog ID, Gaia DR3 IDs, on-sky positions, evolutionary class, core associations of YSOs in the $\rho$ Oph region\label{tab:yso}}
\tablewidth{0pt}
\tabletypesize{\scriptsize}
\tablehead{
\colhead{Source ID} &
\colhead{Gaia DR3 ID} &
\colhead{RA} &
\colhead{Dec} &
\colhead{Class} &
\colhead{Conflict} &
\colhead{SR} &
\colhead{Assoc.} &
\colhead{Core ID} &
\colhead{Core Type}
}
\startdata
SESNA J162715.45$-$242639.8 & \nodata & 246.814393 & $-$24.444393 & II & 0 & 1 & ROPH & \nodata & \nodata \\
eHOPS-oph-25 & \nodata & 246.840960 & $-$24.457696 & F  & 1 & 1 & ROPH & \nodata & \nodata \\
eHOPS-oph-30 & \nodata & 246.866720 & $-$24.659321 & I  & 0 & 1 & ROPH & 162728.1$-$243933 & protostellar \\
SESNA J162224.97$-$232955.2 & 6050204641855013376 & 245.604035 & $-$23.498667 & II & 0 & 1 & ROPH & \nodata & \nodata \\
SESNA J162755.25$-$242839.6 & 6049099460869770240 & 246.980225 & $-$24.477677 & II & 0 & 1 & ROPH & \nodata & \nodata \\
SESNA J162709.08$-$241200.8 & 6049171650679414528 & 246.787817 & $-$24.200232 & II & 0 & 1 & ROPH & \nodata & \nodata \\
eHOPS-oph-11 & \nodata & 246.722810 & $-$24.543443 & F  & 1 & 1 & ROPH & \nodata & \nodata \\
J162546.6$-$242336 & \nodata & 246.444200 & $-$24.393300 & F  & 1 & 1 & ROPH & \nodata & \nodata \\
SESNA J162539.58$-$242635.1 & 6049140142798016384 & 246.414921 & $-$24.443075 & II & 0 & 1 & UPSCO & \nodata & \nodata \\
SESNA J163026.74$-$235908.8 & \nodata & 247.611399 & $-$23.985775 & II & 0 & 1 & ROPH & \nodata & \nodata \\
SESNA J162654.44$-$242620.8 & \nodata & 246.726848 & $-$24.439100 & II & 0 & 1 & ROPH & \nodata & \nodata \\
SESNA J162726.48$-$243923.1 & \nodata & 246.860332 & $-$24.656411 & II & 0 & 1 & ROPH & \nodata & \nodata \\
SESNA J162732.67$-$243323.7 & \nodata & 246.886124 & $-$24.556592 & III & 0 & 1 & ROPH & \nodata & \nodata \\
SESNA J162704.57$-$242715.7 & \nodata & 246.769053 & $-$24.454358 & II & 0 & 1 & ROPH & \nodata & \nodata \\
\enddata

\tablecomments{
The class here is the adopted YSO evolutionary class following our source and class prioritization scheme.
Conflict is a binary flag indicating class conflicts between sources across the catalogs. SR (spatially resolved)
denotes whether the source is spatially resolved when assigning RVs. Core ID is the source ID from the dense core catalog that the YSO got associated with. The raw table provided will also include all IDs of sources that got matched between the catalogs and the classes as each catalog designates them as well as angular separations distances between the matches. The complete version of this table is available and contains 612 members. A portion is shown here for guidance regarding its form and content.}
\end{deluxetable*}

We cross-match our stellar sample with RV data described in section \ref{subsection: kinematic_data}. For each star in our sample, we match it to an RV if both share a \textit{Gaia} ID, or if their associated positions are within 5'' of each other. The full list of sources we draw data from is provided in Table \ref{tab:rv}. We also perform a $5''$ cross-match with {\it Gaia} DR3 for all SESNA, eHOPS, and Dunham sources that do not already have a SPYGLASS counterpart. More than a 100 Gaia sources picked up by the Gaia search did not initially pass our selection criteria within SPYGLASS. We retain these objects, as they are identified in the protostellar catalogs but flag them accordingly.

For stars with Gaia equivalents, we use the \citet{Bailer_2021} distances as a secondary method of checking membership. We use the distance extent of the SPYGLASS members as a reference to define the spatial extent of $\rho$ Oph population. Sources within 129-163 pc are $\rho$ Oph members, those within 163-170 pc are associated with the broader USco population, and sources beyond 170 pc are labeled as non-members. For our analysis focusing on $\rho$ Oph, we retain sources within 163 pc. 

MOCAdb supplemented additional infrared proper motion measurements from \citet{Catwise}, \citet{Ducourant_2017} and that are particularly valuable for the deeply embedded Class 0 or I objects. For these sources, we preferentially adopt the IR derived values whenever available. We further augment the kinematic dataset by assigning $^{13}$CO velocities from the COMPLETE survey to all class 0, I, Flat-spectrum sources, as well as dense cores and as any objects above the 90\% dust column density threshold. This approach assumes that these deeply embedded populations remain coupled to their natal gas and therefore share its line of sight velocity. Table \ref{tab:rv} presents the compiled kinematic data. 


\begin{deluxetable*}{lccccccc}
\tablecaption{Radial velocities, proper motions, distances, and velocity cut flags for $\rho$ Oph members\label{tab:rv}. The full table is available in machine-readable form in the online version of the article.}
\tablewidth{0pt}
\tabletypesize{\scriptsize}
\tablehead{
\colhead{Source ID} &
\colhead{RV} &
\colhead{RV Src \tablenotemark{a}} &
\colhead{$\mu_{\alpha*}$} &
\colhead{$\mu_{\delta}$} &
\colhead{PM Src} &
\colhead{$d$} &
\colhead{Cut}
}
\startdata
SESNA J162715.45$-$242639.8 & -7.09 & 13CO & $-19.40 \pm 4.40$ & $-50.10 \pm 4.40$ & 2021ApJS..253....8M & \nodata & 0 \\
eHOPS-oph-25 & -7.36 & 13CO & $-22.50 \pm 7.30$ & $-50.20 \pm 7.80$ & 2021ApJS..253....8M & \nodata & 0 \\
eHOPS-oph-30 & $-4.37 \pm 0.41$ & Su19 & $-18.40 \pm 5.00$ & $-1.34 \pm 5.20$ & 2021ApJS..253....8M & \nodata & 0 \\
SESNA J162224.97$-$232955.2 & $-5.98 \pm 0.10$ & APOG & $-7.72 \pm 0.05$ & $-26.27 \pm 0.03$ & Gaia DR3 & $136.2^{+0.6}_{-0.6}$ & 1 \\
SESNA J162755.25$-$242839.6 & \nodata & 13CO & $-20.41 \pm 4.40$ & $-49.24 \pm 4.40$ & 2021ApJS..253....8M & \nodata & 0 \\
SESNA J162709.08$-$241200.8 & $-5.94 \pm 0.06$ & AP17 & $-7.00 \pm 0.20$ & $-23.68 \pm 0.14$ & Gaia DR3 & $142.6^{+3.3}_{-2.7}$ & 1 \\
eHOPS-oph-11 & -7.32 & 13CO & $-14.63 \pm 4.50$ & $-51.28 \pm 4.50$ & 2021ApJS..253....8M & \nodata & 0 \\
J162546.6$-$242336 & \nodata & 13CO & \nodata & \nodata & \nodata & \nodata & 0 \\
SESNA J162539.58$-$242635.1 & $-3.66 \pm 0.06$ & AP17 & $-6.52 \pm 0.49$ & $-24.42 \pm 0.37$ & Gaia DR3 & $178.3^{+10.9}_{-12.7}$ & 1 \\
SESNA J163026.74$-$235908.8 & $-7.57 \pm 0.30$ & APOG & \nodata & \nodata & \nodata & \nodata & 0 \\
\enddata

\tablecomments{
The \texttt{Cut} column indicates whether a source did not pass the SPYGLASS velocity-quality criteria.
}
\tablenotetext{a}{The source of the measurement. The codes correspond to the following: Fit to $^{13}CO$ data from the FCRAO COMPLETE Survey \citep{Ridge_2006} ($^{13}$CO), Gaia DR3 (GDR3), new Anglo-Australian Telescope HERMES data (AAT), unreleased SDSS-V APOGEE spectra \citep{APOGEE} (APOG),  GALAH DR4 \citep{Buder_2025} (GAL4), APOGEE DR17 \citep{APOGEEDR17} (AP17), \citet{Muzerolle03} (M3), \citet{Gontcharov06} (G6), \citet{Kurosawa06} (K6), \citet{Torres06} (T6), \citet{Guenther07} (Gu7), \citet{Prato07} (P7), \citet{White07} (W7), \citet{Chen11} (Ch11), \citet{Dahm12} (D12), \citet{Galli13} (G13), \citet{Kuruwita18} (K18), \citet{Sullivan19} (Su19), \citet{GutierrezAlbarran20} (GA20), \citet{MiretRoig22} (MR22), \citep{Randich_2022} (GES)}
\end{deluxetable*}

\subsection{Clustering}
\label{subsection:clusters}
To identify spatial and kinematic substructures within the $\rho$~Oph region, we applied the \texttt{HDBSCAN} algorithm \citep{HDBSCAN} to the YSO and dense core datasets. The clustering was performed in a three-dimensional parameter space defined by RA, Dec, and RV. We choose RV as the kinematic dimension because it allows to include the dense cores and YSOs in the clustering. Prior to clustering, each dimension was standardized using \texttt{StandardScaler} from \texttt{scikit-learn}, which centers each feature to zero mean and scales it to unit variance, placing the three coordinates on comparable numerical scales. Dense cores with associated protostars that have measured RVs were excluded from the clustering analysis to avoid any redundant representation. We adopted \texttt{min\_cluster\_size = 6} and \texttt{min\_samples = 6}, with \texttt{cluster\_selection\_method = ``eom''} \citep{scikit}. The clustering recovered major subregions of $\rho$ Oph complex such as L1688, L1689A/B, and L1709, as well as two populations north of L1688, one of which is associated with the $\rho$ Oph star, and older populations to the west consisting of SPYGLASS sources. Any sources that were excluded from the clustering due to missing RVs were reassigned to the nearest on-sky cluster center. 

\begin{figure*}[t]
    \centering
    \includegraphics[width=0.99\textwidth]{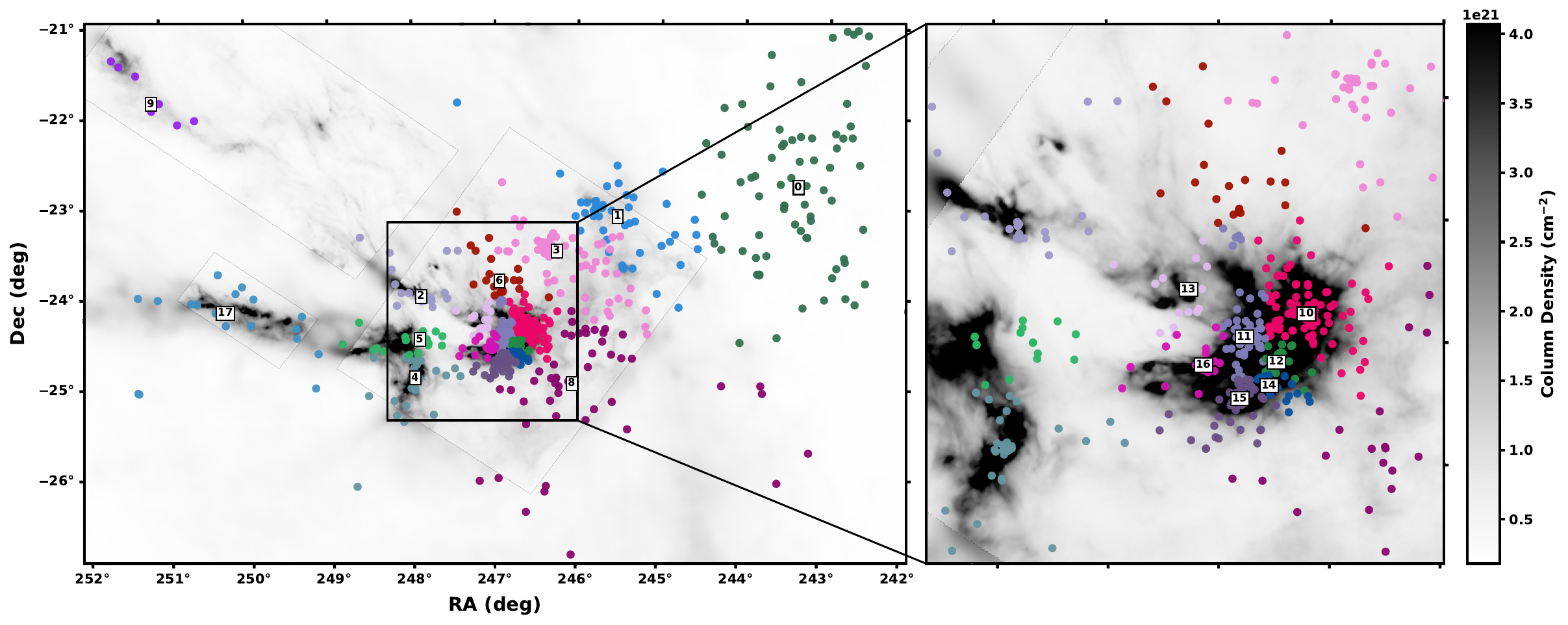}
    \caption{The spatial clustering of sources in $\rho$ Oph. The left panel shows the all the sources, with subregions labeled by their HDBSCAN cluster assignments derived from their (RA, Dec, and RV). The right panel shows the zoomed-in view of the clustered source distribution in the main L1688, L1709 and L1689A/B regions}. 
    \label{fig:clusters}
\end{figure*}

The $\rho$ Oph cloud contains several well-known substructures including the Oph A-G dense core regions identified in \citet{Loren_1989}. Within L1688, we group these components spatially, as illustrated in Fig. \ref{fig:clusters}. These regions are defined using bounding boxes in RA-Dec, expanded by a buffer radius of 15' beyond the original boundary. This buffer radius allows the off-cloud population to extend further from the embedded cluster, better separating the tenuous, diffuse off-cloud population from the dense on-cloud population.  We also separated the eastern extensions (L1729 and L1712) identified by \citet{Loren_1989}, and a distinct northwestern streamer (Cluster 11) that is spatially separated from the main cloud and coincident with a streamer in $^{13}$CO also identified in \citet{Loren_1989}. These sources were excluded from the HDBSCAN clusters due to a lack of RV measurements. We summarize the cluster designations, the census of YSOs by evolutionary class, and the number of associated dense cores in each region in table \ref{tab:cluster_class}. 

To assess the robustness of this decomposition, we repeated the clustering over 30 parameter combinations, with \texttt{min\_cluster\_size} {3,4,5,6,8,10,12,15} and \texttt{min\_samples} {3,4,6,8,10,12,15} (\texttt{min\_samples}  \texttt{min\_cluster\_size}). This range runs from the dimensionality of the data to values large enough that our smallest clusters can no longer form. For each star, we computed the fraction of runs in which it is grouped with the other members of its adopted cluster, considering only runs in which a cluster of that size is permitted. The mean co-clustering fractions are 0.84 for L1688, 0.99 for L1709, 0.83 for L1689A, 0.76 for Cluster 0, 0.69 for Cluster 1, 0.68 for Cluster 3, and and 1.00 for Cluster 6. 

To assess the influence of the gas-derived velocities, we also repeated the clustering using only sources with measured stellar RVs, excluding both the dense cores and the YSOs whose RVs are $^{13}$CO proxies. This recovers Clusters 0, 1 and 3, L1688 and L1689. Repeating this clustering over the same parameter grid gives mean co-clustering fractions of 0.71 for Cluster 0, 0.67 for Cluster 1, 0.66 for Cluster 3, 0.70 for L1688 and 0.98 for L1689.

\begin{deluxetable*}{lccccccccccccccc}
\tablecaption{Distribution of the YSO classes across clusters in $\rho$ Oph region, the isochronal ages for selected clusters, and the protostar derived ages for each of the clusters \label{tab:cluster_class}}
\tablewidth{0pt}
\tabletypesize{\scriptsize}
\tablehead{
\colhead{Cluster} & 
\colhead{0} &
\colhead{I} &
\colhead{F} &
\colhead{II} &
\colhead{III} &
\colhead{PMS} &
\colhead{Redd.\ Disk} &
\colhead{Disk} &
\colhead{Core} & 
\colhead{$(0+\mathrm{I})/\mathrm{II}$} &
\colhead{$(0+\mathrm{I}+\mathrm{F})/\mathrm{II}$} &
\colhead{Isochronal Age (Myr)} &
\colhead{Final Age (Myr)} &
\colhead{On-cloud Age (Myr)} &
\colhead{Off-cloud Age (Myr)}
}
\startdata
0  & 0 & 0  & 0  & 0  & 0 & 74 & 0 & 0 & 0  & \nodata & \nodata & $7.94^{+3.42}_{-3.41}$ & $7.94^{+3.42}_{-3.41}$ & \nodata & $7.94^{+3.42}_{-3.41}$\\
1  & 0 & 1  & 4  & 12 & 3 & 24 & 0 & 0 & 0  & 0.08 & 0.42 & $3.80^{+1.88}_{-1.78}$ & $2.95^{+1.15}_{-1.09}$ & \nodata & $3.00^{+1.17}_{-1.11}$\\
2  & 2 & 6  & 3  & 7  & 3 & 2  & 0 & 0 & 9  & 1.14 & 1.57 & \nodata & $0.61^{+0.07}_{-0.07}$ & $0.50^{+0.09}_{-0.08}$ & $0.77^{+0.11}_{-0.11}$\\
3  & 0 & 4  & 1  & 17 & 0 & 39 & 0 & 1 & 1  & 0.24 & 0.29 & $3.63^{+1.62}_{-1.56}$ & $2.86^{+1.02}_{-0.97}$ & \nodata & $2.86^{+1.02}_{-0.97}$\\
4  & 1 & 7  & 0  & 22 & 0 & 2  & 0 & 1 & 28 & 0.36 & 0.36 & \nodata & $0.68^{+0.06}_{-0.06}$ & \nodata & $1.51^{+0.13}_{-0.13}$\\
5  & 3 & 2  & 0  & 10 & 1 & 3  & 0 & 2 & 6  & 0.50 & 0.50 & \nodata & $0.92^{+0.10}_{-0.10}$ & $0.26^{+0.09}_{-0.10}$ & $1.51^{+0.16}_{-0.16}$\\
6  & 0 & 2  & 0  & 8  & 0 & 9  & 0 & 3 & 0  & 0.25 & 0.25 & $4.17^{+1.94}_{-2.09}$ & $2.83^{+0.93}_{-0.97}$ & \nodata & $2.83^{+0.93}_{-0.97}$\\
8  & 0 & 5  & 1  & 11 & 0 & 29 & 0 & 1 & 0  & 0.45 & 0.55 & $5.24^{+2.34}_{-2.31}$ & $3.84^{+1.48}_{-1.47}$ & \nodata & $3.84^{+1.48}_{-1.47}$\\
9  & 0 & 5  & 0  & 2  & 0 & 0  & 0 & 0 & 0  & 2.50 & 2.50 & \nodata & $0.84^{+0.14}_{-0.13}$ & $1.18^{+0.23}_{-0.22}$ & $0.39^{+0.08}_{-0.07}$\\
10 & 2 & 10 & 13 & 50 & 3 & 9  & 1 & 1 & 38 & 0.24 & 0.50 & $2.09^{+1.28}_{-1.03}$ & $1.03^{+0.10}_{-0.10}$ & $0.87^{+0.09}_{-0.09}$ & $1.79^{+0.21}_{-0.21}$\\
11 & 0 & 3  & 10 & 31 & 4 & 3  & 0 & 0 & 21 & 0.10 & 0.42 & \nodata & $1.00^{+0.07}_{-0.08}$ & $0.90^{+0.07}_{-0.07}$ & $1.97^{+0.25}_{-0.26}$\\
12 & 2 & 4  & 3  & 10 & 1 & 0  & 1 & 1 & 4  & 0.60 & 0.90 & \nodata & $0.99^{+0.10}_{-0.10}$ & $0.95^{+0.10}_{-0.09}$ & \nodata\\
13 & 0 & 2  & 0  & 8  & 3 & 2  & 0 & 0 & 5  & 0.25 & 0.25 & \nodata & $1.02^{+0.12}_{-0.12}$ & $0.12^{+0.12}_{-0.12}$ & $1.62^{+0.19}_{-0.19}$\\
14 & 0 & 2  & 3  & 13 & 0 & 5  & 0 & 0 & 5  & 0.15 & 0.38 & \nodata & $1.19^{+0.11}_{-0.11}$ & $1.20^{+0.12}_{-0.11}$ & $1.18^{+0.33}_{-0.34}$\\
15 & 3 & 7  & 11 & 15 & 1 & 6  & 1 & 1 & 5  & 0.67 & 1.40 & \nodata & $0.97^{+0.08}_{-0.08}$ & $0.79^{+0.07}_{-0.07}$ & $1.73^{+0.21}_{-0.21}$\\
16 & 1 & 1  & 0  & 10 & 2 & 1  & 0 & 0 & 8  & 0.20 & 0.20 & \nodata & $0.91^{+0.10}_{-0.10}$ & $0.55^{+0.10}_{-0.09}$ & $1.97^{+0.27}_{-0.28}$\\
17 & 0 & 6  & 0  & 6  & 1 & 4  & 0 & 2 & 0  & 1.00 & 1.00 & \nodata & $1.18^{+0.13}_{-0.13}$ & $1.18^{+0.34}_{-0.34}$ & $1.18^{+0.14}_{-0.14}$\\
\enddata
\tablecomments{Cluster 12 has a single off-cloud source, we do not report the age for off cloud here.}
\end{deluxetable*}

\subsection{Compression Arc Decomposition}
\label{subsection: arcs}

\citet{Ladjelate_2020} identify three distinct compression arcs within L1688, and propose that these arc-like structures may originate from multiple successive shock-waves propagating through the cloud from the direction of the SCO OB2. To test this picture and investigate whether an age gradient or kinematic signature is present between these compression fronts, we adopt an alternate spatial grouping of $\rho$ Oph based on these arcs. We adopt their Arc 1, roughly comprising Oph A, C, E + F, and merge their Arc 2 and 3 into a single region (hereafter Arc 2), comprising Oph G and B. Our assignment of already-resolved subgroups to these arcs is approximate rather than identical to that of \citet{Ladjelate_2020}, but provides a framework for assessing both the age sequence across the compression fronts and the relative kinematics of populations lying along them. 

\subsection{On-Cloud and Off Cloud Decomposition}
\label{subsection: on_off}

\begin{figure}[t]
    \centering
\includegraphics[width=\columnwidth]{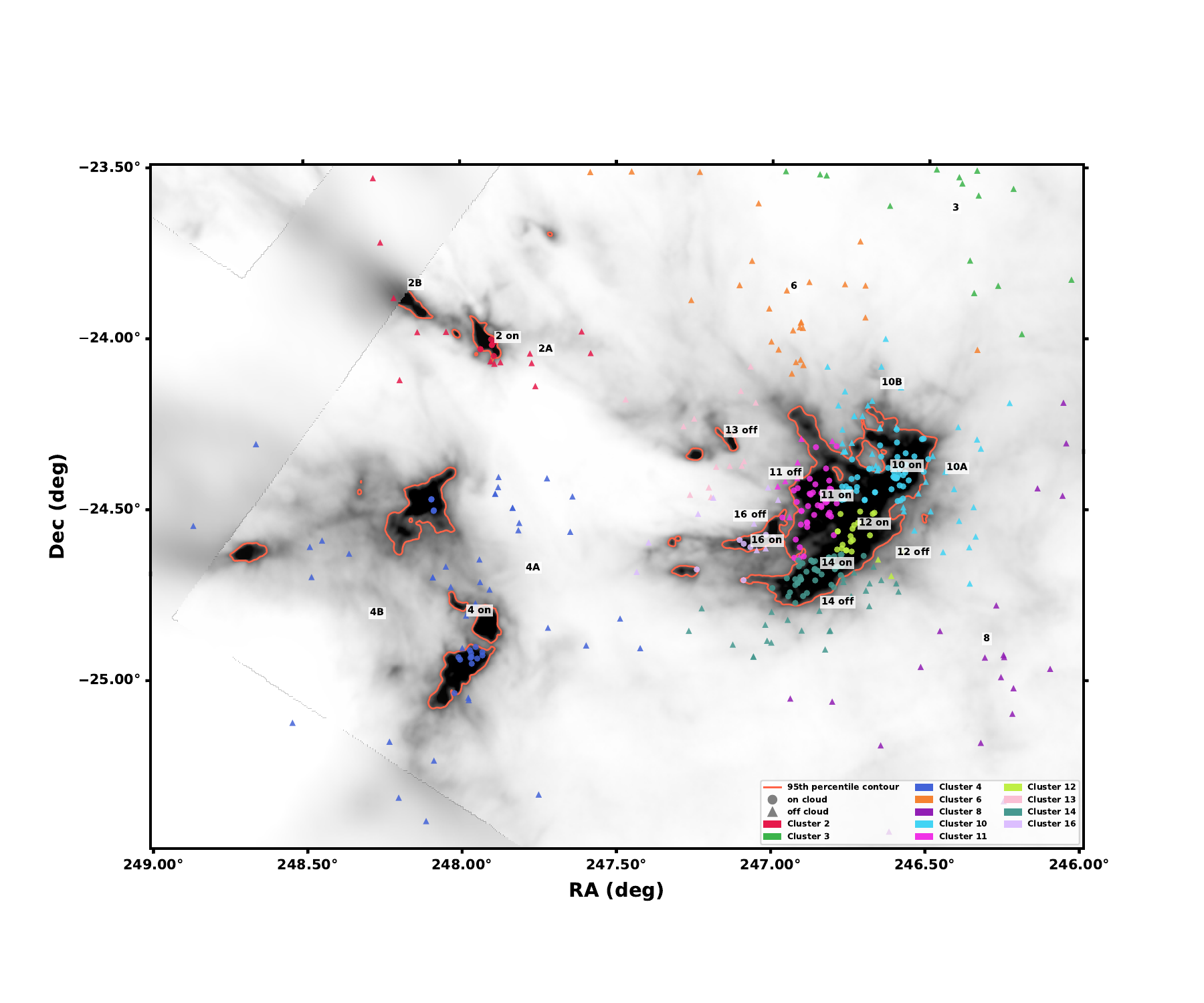}
    \caption{Spatial distribution of subregions, their on and off cloud components, the 95th column density threshold percentile for reference.}
    \label{fig: offon}
\end{figure}

Several subregions within $\rho$ Oph show spatially distinct stellar populations offset from their associated dense gas, in some cases distributed across multiple directions on the sky. If large scale winds or radiation from SCO OB2 largely from the southwestern direction was driving cloud acceleration \citet{Ladjelate_2020}, we would expect to observe stars kinematically decoupled from the dense gas and displaced in the upwind direction. In order to identify potential localized and global feedback signatures, we separate each relevant subregion into their on- and off-cloud components. Sources are classified as on cloud if they lie within the 95th percentile contour of the Herschel column density maps. Some subregions have off-cloud sources distributed across multiple spatial directions, which we treat as separate components in our kinematic analysis. We split up the populations in RA-Dec. The resulting on and off cloud classifications are shown in Figure \ref{fig: offon}. 

\subsection{Ages} 
\label{subsection:ages}
In this section we estimate the ages of subregions within $\rho$ Oph by combining two complementary age determining methods. For the younger, embedded population, we leverage  protostellar class / prestellar core lifetimes. For the more evolved, optically visible pre-main-sequence population, which lacks a well-defined lifetime, we instead use isochronal ages. Together, the two methods allow for the construction of a holistic regional age estimate encompassing the full YSO and dense core population. 

YSO classes are commonly used as statistical age tracers. Their lifetimes can be inferred from the relative counts of sources in each class under the assumption of continuous star formation \citep{Evans_2009, Dunham_2015, Wilking_1989, Greene_1994}. We estimate cluster ages by assigning a class-based characteristic age to each source based on its evolutionary class, under the assumption of a monotonic progression through the sequence: prestellar core, Class 0, Class I, Flat-spectrum, Class II, Class III/SPYGLASS sources. We define the zero point of this sequence as the onset of protostellar collapse, corresponding to the transition from prestellar core to Class 0. For the protostellar classes, we adopt the lifetimes from \citet{Dunham_2015}, who estimate class durations from population statistics across nearby star forming regions: $0.15$--$0.24$ Myr for Class 0, $0.31$--$0.48$ Myr for Class I, $0.26$--$0.50$ Myr for Flat-spectrum sources, and a fixed duration of 2 Myr to the Class II phase.

We derive a characteristic age and uncertainty for each class using a Monte Carlo approach with $10^5$ realizations. In each realization, we draw the lifetime of each class uniformly from its range (the Class II lifetime is fixed at 2.0\,Myr), and place the classes in sequence so that each begins where the previous one ends. We then draw an age uniformly from within each class interval, as expected for a roughly constant star formation rate. The median and 16th--84th percentiles of the resulting distributions give characteristic ages of $0.10 \pm 0.07$\,Myr for Class 0, $0.39^{+0.14}_{-0.13}$\,Myr for Class I, $0.77^{+0.14}_{-0.13}$\,Myr for Flat-spectrum sources and $1.97 \pm 0.68$\,Myr for Class II sources. These uncertainties reflect both the uncertainty in the class lifetimes and the spread of ages within each class. Prestellar cores are assigned a lifetime of 0.5\,Myr \citep{Evans_core_2009} preceding the onset of Class 0, giving a characteristic age of $-0.25 \pm 0.17$\,Myr. 

\begin{figure*}[t]
    \centering
    \includegraphics[width=0.7\textwidth]{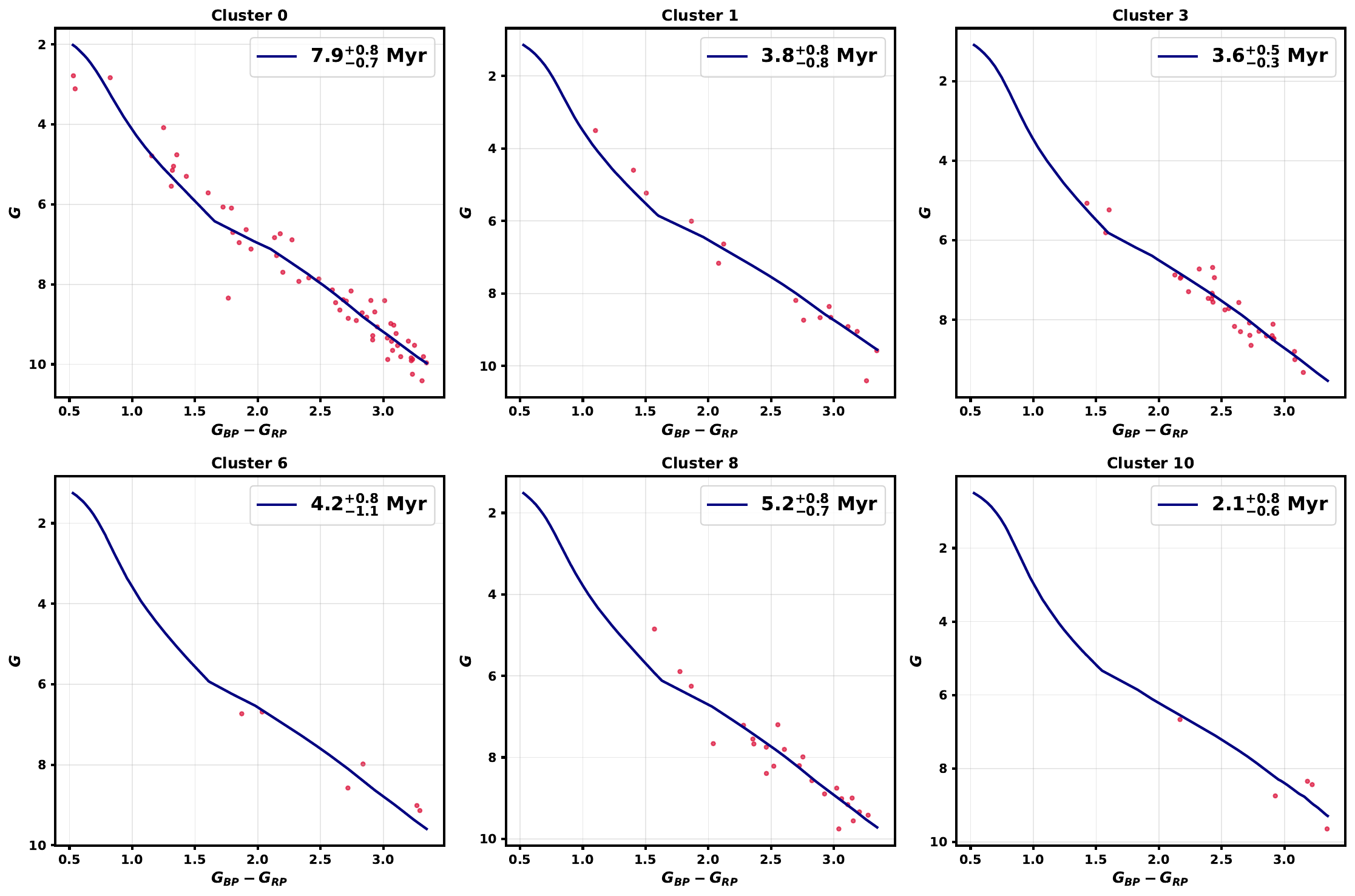}
    \caption{Isochronal fits for clusters containing more than five SPYGLASS sources. Only sources without protostellar counterparts are included in the fits. The best fitting PARSEC Isochrones are overlaid on the Colour-Magnitude Diagram for each cluster. The star $\rho$ Oph is in the region nearby Cluster 3}
    \label{fig:spyglass_isochrone}
\end{figure*}

To estimate representative isochronal ages for the Class III and later population in $\rho$ Oph, we use the PARSEC v1.2S Isochrones models \citep{Chen_2015}, which are well suited for the assortment of more evolved, non-embedded pre-main-sequence stars that SPYGLASS primarily traces. We determine a age for each cluster defined in Section \ref{subsection:clusters} containing more than give associated SPYGLASS stars. SPYGLASS corrects each star's photometry for extinction individually, using the extinction to its distance from the dust maps of \citet{Lallement_2019}. We compute ages using least squares fit between extinction corrected $G$-band magnitudes and PARSEC model over a color-magnitude grid spanning the colour range common to all the isochrones in each model, with reddening corrections from \citet{Wang_2019}, excluding stars deviating more than 3$\sigma$ from the isochrone in G-band magnitude prior to fitting. Since most SPYGLASS stars are not embedded, the limited resolution of the maps in the densest parts of the cloud should have little effect on the isochronal sequence. The best fit isochronal ages for the relevant clusters are shown in Table \ref{tab:cluster_class}, and we show the resulting isochronal fits for six of the clusters in Fig. \ref{fig:spyglass_isochrone}. We estimate age uncertainties through through bootstrap resampling, in which cluster members are resampled  with replacement 1000 times and isochrone fit repeated each time. Asymmetric 1$\sigma$ uncertainties are given by the 16th and 84th percentile of the age distribution. The bootstrap uncertainties capture only the statistical uncertainty within the adopted PARSEC framework and not systematic uncertainties. To estimate the systematic uncertainty from the choice of evolutionary model, we use the ages of Upper Sco derived from eight model grids by \citet{Herczeg_2015}. These have a mean of 4.9\,Myr and a standard deviation of 2.1\,Myr, a fractional spread of 42\%. We therefore adopt a systematic uncertainty of $0.42\times$ the age, and report each isochronal age as $t \pm \sigma_{\mathrm{stat}} \pm \sigma_{\mathrm{sys}}$. For each cluster, we compute the mean age by averaging the assigned ages of all constituent sources both protostellar and non protostellar. Uncertainties on the cluster mean ages are estimated by combining the standard error on the mean with the propagated uncertainties on individual source age assignment.  

To test whether the cluster ages depend on the treatment of Flat-spectrum sources, we recomputed the mean age of each cluster under two alternative assumptions: Flat sources assigned the Class I or the Class II age. For every cluster, the baseline mean age lies between these two extremes. Eight of the 17 clusters contain no Flat sources and are therefore unchanged. A further four (clusters 1, 3, 4 and 8) change by less than 4\% ($\leq 0.11$\,Myr). The remaining six (clusters 2, 10, 11, 12, 14 and 15) contain a larger fraction of Flat sources and shift by 12--30\% (0.13--0.30\,Myr), with the largest change in cluster 15, whose age of 1.02\,Myr becomes 0.92\,Myr when Flat sources are treated as protostars and 1.32\,Myr when they are treated as Class II sources.

We also tested whether the on-cloud and off-cloud ages depend on the column-density threshold used to define the cloud. We repeated the classification using the 80th, 85th, 90th and 95th percentiles of the Herschel column density. At every threshold the off-cloud populations are older than the on-cloud ones in most clusters that contain both groups, with a mean off-cloud minus on-cloud difference of +0.7 to +0.9\,Myr (median 1\,Myr). For the seven clusters that contain both groups at every threshold, the mean difference changes by only 0.15\,Myr between the 80th and 95th percentiles (from 0.77 to +0.88\,Myr).

\section{Results}

\subsection{Age Patterns}
\begin{figure}[t]
    \centering
\includegraphics[width=\columnwidth]{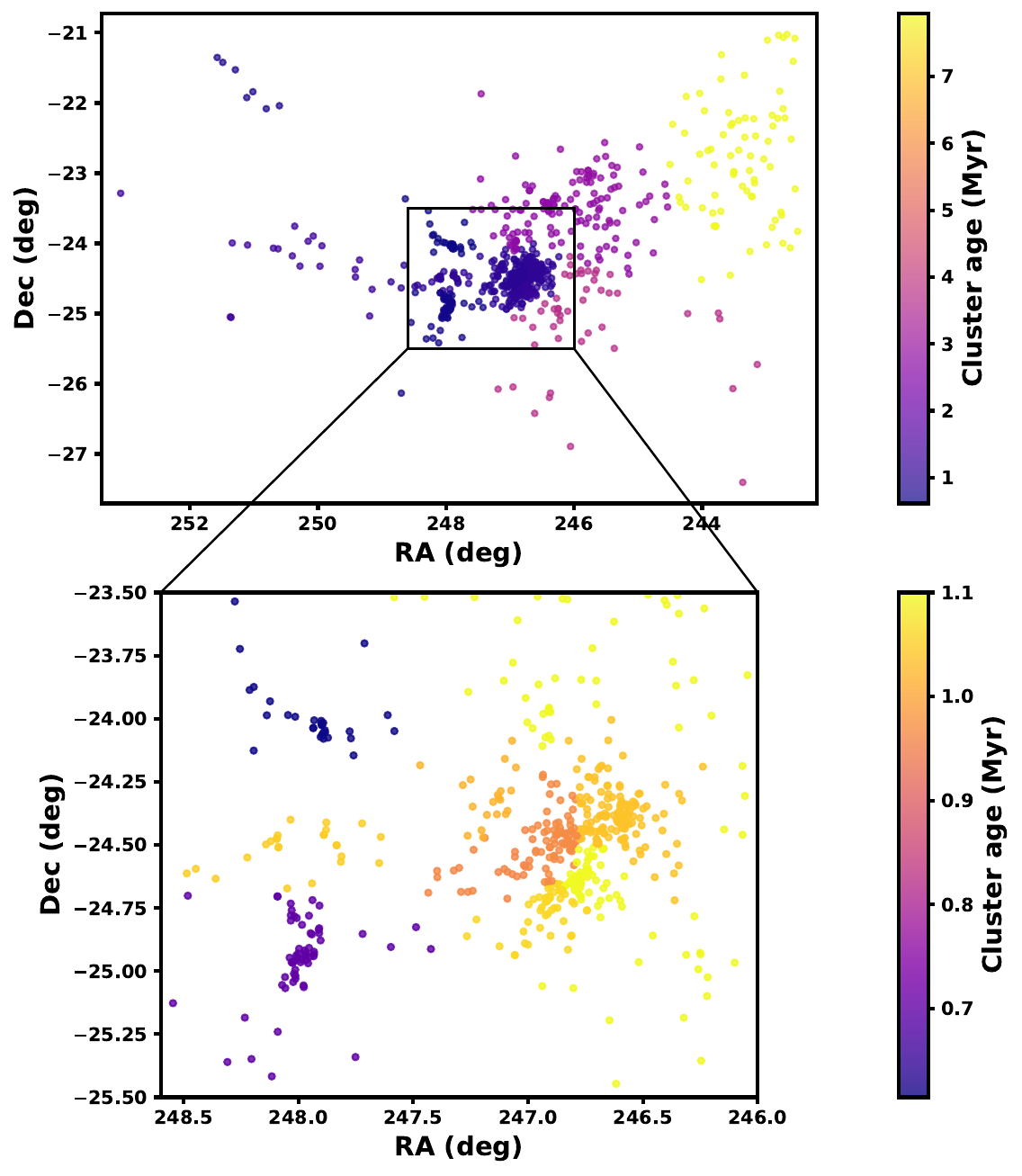}
    \caption{Spatial distribution of the protostellar derived ages in the $\rho$ Oph region. In the top panel, shows all the identified subgroups coloured by their derived ages. The bottom panel zooms in on L1688 and adjacent regions, accompanied by an adjusted colour scale to better show the age gradients present. The oldest population is Cluster 0. }
    \label{fig:prot_age}
\end{figure}

Figure \ref{fig:prot_age} shows the on-sky distribution of sources within each subregion of $\rho$ Oph, coloured by the age assigned via the combined protostellar class-based and isochronal age estimation methods described in section \ref{subsection:ages}.  A broad age gradient from west to east is apparent, with the oldest population in cluster 0 giving way to progressively younger subregions towards the core $\rho$ Oph cloud. Locally, the peripheral $\rho$ Oph subregions (Oph A, C, E, F) are marginally older than Oph B, D, and G,  although the differences are comparable to the statistical uncertainties. The youngest populations, however, are found in L1709 and L1689 to the east, alongside L1712 and L1755. Stellar populations adjacent to L1688, such as Cluster 8, are older than the cloud itself.

Within our compression arc framework, we find that Arc 1's on-cloud population has a mean age of $0.85 \pm 0.09$\,Myr and the Arc 2 on-cloud population has $0.74 \pm 0.13$\,Myr. We note that the age difference between the two arcs is also sensitive to the off-cloud components lying to the lower-right of L1688, which are systematically older and consequently pull up the mean age of their associated subregions however the ages are indistinguishable given their uncertainties.

Our age solution for the on- and off-cloud samples reveal a consistent and expected pattern: the off cloud populations are systematically older than the on-cloud population by a typical offset of $\sim 1.0$ Myr. The largest on/off-cloud age difference is in Oph G ($\Delta$age $\sim 1.4$\,Myr), although its off-cloud component contains only 8 stars and this difference should be regarded as indicative. A notable exception is Oph E, which shows nearly identical on- and off-cloud ages. Compared to the other Arc 1 subregions, most notably Oph A and Oph F, Oph E shows less spatial separation between its on- and off-cloud components, which likely contributes to the near-zero age difference observed for this subregion (Cluster 14). Oph D has the youngest on-cloud characteristic age, driven largely by its high fraction of dense cores. For L1709 and L1689A/B (cluster 2, 4, 5), the on cloud populations are substantially younger than their off cloud counterparts and main L1688 subregions. The off cloud populations of L1709 and L1689 are younger than typical L1688 off cloud ages. 


\subsection{Kinematics}
\label{sec: on and off cloud}

Feedback-driven star formation, cloud compression and decoupling of stellar populations from their birth sites produce distinct kinematic signatures. For each subregion defined in section \ref{subsection: on_off}, we clip on both proper motion components and RV, removing sources that deviate more than 2$\sigma$ from the median to remove sources with discrepant kinematics. We then compute the perspective corrected transverse velocities following \citet{Kuhn_2019} and estimate uncertainties of the bulk velocities by performing a bootstrap resampling analysis prior to clipping on each subgroup. We present the bulk perspective-corrected transverse velocity for each subgroup in Table \ref{tab:vtlvtb}. 

\begin{deluxetable*}{lcccccc}
\tablecaption{Mean perspective-corrected transverse velocities, bootstrap uncertainties, and bulk heliocentric UVW velocities for each subregion and on/off cloud population of $\rho$ Oph. \label{tab:vtlvtb}}
\tablewidth{0pt}
\tabletypesize{\scriptsize}
\tablehead{
\colhead{Subgroup} &
\colhead{$N$} &
\colhead{$\bar{v}_{tl}$ (km s$^{-1}$)} &
\colhead{$\bar{v}_{tb}$ (km s$^{-1}$)} &
\colhead{$U$ (km s$^{-1}$)} &
\colhead{$V$ (km s$^{-1}$)} &
\colhead{$W$ (km s$^{-1}$)}
}
\startdata
0 \ldots & 55 & $-17.54 \pm 0.14$ & $-4.93 \pm 0.20$ & $-4.88 \pm 0.27$ & $-15.85 \pm 0.08$ & $-8.09 \pm 0.13$ \\
\tableline
1 \ldots & 21 & $-17.40 \pm 0.21$ & $-5.02 \pm 0.37$ & $-6.17 \pm 0.34$ & $-15.39 \pm 0.11$ & $-8.93 \pm 0.17$ \\
\tableline
2 on (L1709 on cloud)            &  3 & $-16.04 \pm 0.46$ & $-5.87 \pm 0.09$ & $-6.18 \pm 0.18$ & $-14.10 \pm 0.64$ & $-9.41 \pm 0.27$ \\
2A (L1709 West off cloud)        &  2 & $-16.37 \pm 0.18$ & $-5.83 \pm 0.18$ & $-6.31 \pm 0.21$ & $-14.39 \pm 0.21$ & $-9.44 \pm 0.23$ \\
2B (L1709 East off cloud)        &  1 & $-16.93$          & $-6.72$          & $-5.40$          & $-15.23$& $-9.85$          \\
\tableline
3 \ldots & 39 & $-17.01 \pm 0.11$ & $-5.87 \pm 0.17$ & $-6.02 \pm 0.16$ & $-15.01 \pm 0.09$ & $-9.60 \pm 0.09$ \\
\tableline
4 on (L1689 on cloud) &  3 & $-17.40 \pm 1.76$ & $-6.07 \pm 2.24$ & $-5.30 \pm 0.06$ & $-16.18 \pm 0.95$ & $-8.20 \pm 0.09$ \\
4A (L1689 West off cloud)        &  8 & $-15.86 \pm 0.34$ & $-6.47 \pm 0.25$ & $-3.83 \pm 0.89$ & $-14.63 \pm 0.10$ & $-8.54 \pm 0.35$ \\
4B (L1689 East off cloud)        &  4 & $-15.91 \pm 0.22$ & $-6.07 \pm 0.25$ & $-5.22 \pm 0.23$ & $-14.26 \pm 0.27$ & $-8.75 \pm 0.40$ \\
\tableline
6 (North of L1688)               & 14 & $-17.25 \pm 0.12$ & $-6.31 \pm 0.24$ & $-5.90 \pm 0.19$ & $-15.32 \pm 0.12$ & $-9.83 \pm 0.15$ \\
\tableline
8 \ldots                      & 21 & $-17.13 \pm 0.31$ & $-6.43 \pm 0.38$ & $-4.70 \pm 0.26$ & $-15.51 \pm 0.19$ & $-8.80 \pm 0.21$ \\
\tableline
10 on (Oph A on cloud)           &  9 & $-17.29 \pm 0.59$ & $-6.01 \pm 0.65$ & $-5.72 \pm 0.26$ & $-15.05 \pm 0.20$ & $-8.78 \pm 0.18$ \\
10A (Oph A North off cloud)      &  5 & $-17.36 \pm 0.65$ & $-7.34 \pm 0.37$ & $-4.70 \pm 0.50$ & $-15.08 \pm 0.28$ & $-9.76 \pm 0.40$ \\
10B (Oph A West off cloud)       &  3 & $-17.51 \pm 0.68$ & $-6.38 \pm 0.68$ & $-5.95 \pm 0.46$ & $-15.53 \pm 0.43$ & $-9.85 \pm 0.75$ \\
\tableline
11 on (Oph B on cloud)           &  9 & $-18.94 \pm 0.45$ & $-6.75 \pm 0.64$ & $-6.27 \pm 0.12$ & $-16.86 \pm 0.57$ & $-10.24 \pm 0.49$ \\
11 off (Oph B off cloud)         &  2 & $-17.71 \pm 0.20$ & $-5.13 \pm 0.27$ & $-6.25 \pm 0.04$ & $-15.71 \pm 0.25$ & $-8.49 \pm 0.45$ \\
\tableline
12 on (Oph C on cloud)           &  3 & $-18.05 \pm 0.71$ & $-7.16 \pm 0.70$ & $-5.92 \pm 0.35$ & $-15.74 \pm 0.98$ & $-10.23 \pm 0.90$ \\
\tableline
13 off (Oph D off cloud)         &  7 & $-17.45 \pm 0.24$ & $-5.15 \pm 0.46$ & $-7.68 \pm 0.77$ & $-14.96 \pm 0.30$ & $-9.46 \pm 0.28$ \\
\tableline
14 on (Oph E + F on cloud)       &  6 & $-17.33 \pm 0.43$ & $-6.82 \pm 0.28$ & $-5.49 \pm 0.34$ & $-14.95 \pm 0.27$ & $-9.45 \pm 0.30$ \\
14 off (Oph E + F off cloud)     & 11 & $-17.29 \pm 0.30$ & $-6.17 \pm 0.24$ & $-6.04 \pm 0.30$ & $-15.24 \pm 0.25$ & $-9.45 \pm 0.32$ \\
\tableline
16 on (Oph G on cloud)           &  3 & $-19.61 \pm 1.92$ & $-6.98 \pm 0.50$ & $-5.91 \pm 0.32$ & $-17.68 \pm 2.25$ & $-10.16 \pm 0.63$ \\
16 off (Oph G off cloud)         &  2 & $-16.86 \pm 0.23$ & $-5.97 \pm 0.38$ & $-4.83 \pm 0.15$ & $-15.30 \pm 0.39$ & $-8.56 \pm 0.33$ \\
\tableline
17 \ldots                     &  2 & $-15.01 \pm 0.29$ & $-6.97 \pm 0.16$ & $-3.47 \pm 1.60$ & $-14.03 \pm 0.11$ & $-8.62 \pm 0.35$ \\
\enddata
\tablecomments{Velocities are perspective-corrected transverse velocities in Galactic coordinates ($v_{tl}$, $v_{tb}$), and bulk heliocentric UVW velocities. $N$ is the number of sources used in the $v_{tl}$/$v_{tb}$ computation after $2\sigma$ clipping on proper motions and radial velocities. Uncertainties on $v_{tl}$/$v_{tb}$ are bootstrap standard errors from 1000 resamplings; uncertainties on UVW are the standard error of the mean. UVW is computed separately for each listed subgroup using the same $2\sigma$ clipping procedure. For subgroups with only one retained source, no uncertainty can be estimated. Cluster 8 is a manually defined structure adjacent to L1688 and was excluded from the UVW comparison since it does not correspond to a resolved on/off-cloud subgroup.}
\end{deluxetable*}

\begin{figure}[t]
    \centering
\includegraphics[width=\columnwidth]{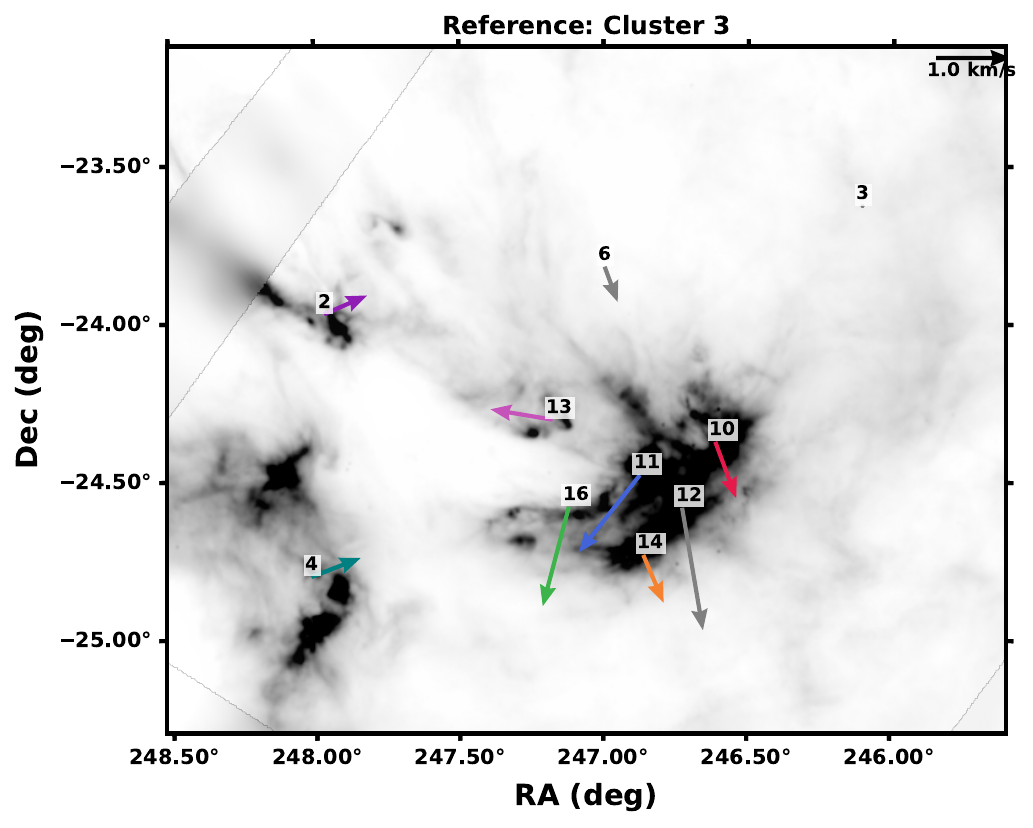}
    \caption{Separation velocity of the main L1688, L1709, and L1689 subgroups from the older, more dispersed cluster 3. Each vector represents the difference in perspective-corrected bulk transverse velocity between the subgroup and the older reference population, rooted at the median position of the subgroup.}
    \label{fig: subgroup relative to 0&1}
\end{figure}

Figure \ref{fig: subgroup relative to 0&1} shows the velocities of the main L1688, L1709, and L1689 subgroups relative to the older, more dispersed subregion (Clusters 3) to the north of $\rho$ Oph. The main L1688 subgroups (10, 11, 12, 14, 16, 13, and 15) differ from the older subgroups by $\sim$ 1 km/s. Within L1688, velocities across L1688 produce a directional dichotomy: subgroups along the western edge of the cloud (Oph A, C, E + F) (Arc 1) show a slight westward component in their separation velocities, while Oph B and G  (Arc 2) move more southeastward. L1709, L1689, and Oph D, by contrast do not exhibit the same coherent separation away from the older northern subgroups seen in the main L1688 subregions and appear kinematically distinct. To investigate the origin of this, we examine the kinematics of $\rho$ Oph at progressively finer scales in the following subsections.

\subsubsection{Velocities along the Arcs}
\label{sec:arcs}
The compression arcs identified by \citet{Ladjelate_2020} provide a useful way to investigate whether kinematic signatures persist at finer scales and whether relative motions between arcs are consistent with ongoing feedback-driven compression from Sco OB2. Figure 8 presents the velocity differences between the on-cloud components of Arc 1 (Oph A, Oph C, and Oph E + F) and the off-cloud components of Oph A, Oph E + F, and Group 8. It also shows the velocity differences between Arc 2 (Oph B and Oph G) and Group 8. Arc 1 shows a separation velocity of $\Delta v_{tl} = -0.34 \pm 0.24$\,km\,s$^{-1}$ and $\Delta v_{tb} = 0.21 \pm 0.24$\,km\,s$^{-1}$. Arc 2 shows a larger, southeastward velocity relative to Group 8, with $\Delta v_{tl} = -1.15 \pm 0.53$\,km\,s$^{-1}$ and $\Delta v_{tb} = 0.23 \pm 0.38$\,km\,s$^{-1}$. Oph D shows a northward motion with a slight eastward component relative to Group 8, broadly aligned with the axis of the local streamer, with $\Delta v_{tl} = -0.32 \pm 0.32$\,km\,s$^{-1}$ and $\Delta v_{tb} = 1.28 \pm 0.41$\,km\,s$^{-1}$.

To assess whether the two arcs move differently, we compute the difference in their mean perspective-corrected transverse velocities, $\Delta\boldsymbol{v} = \bar{\boldsymbol{v}}_{\rm Arc\,1} - \bar{\boldsymbol{v}}_{\rm Arc\,2}$, with uncertainties from bootstrap resampling of the members of each arc. We test whether $\Delta\boldsymbol{v}$ is consistent with zero using a permutation test, in which arc membership is randomly reassigned among the combined sample and a covariance-weighted measure of $\Delta\boldsymbol{v}$ is recomputed. Between Arc 1 ($N = 45$) and Arc 2 ($N = 15$), we find $\Delta v_{tl} = 0.78 \pm 0.25$\,km\,s$^{-1}$ and $\Delta v_{tb} = 0.01 \pm 0.28$\,km\,s$^{-1}$, an offset of $3.2\sigma$ from zero in the bootstrap distribution (permutation $p = 0.02$). The arcs therefore show moderate evidence for a difference in mean transverse velocity, confined to the $v_{tl}$ direction.

\begin{figure}[t]
    \centering
\includegraphics[width=\columnwidth]{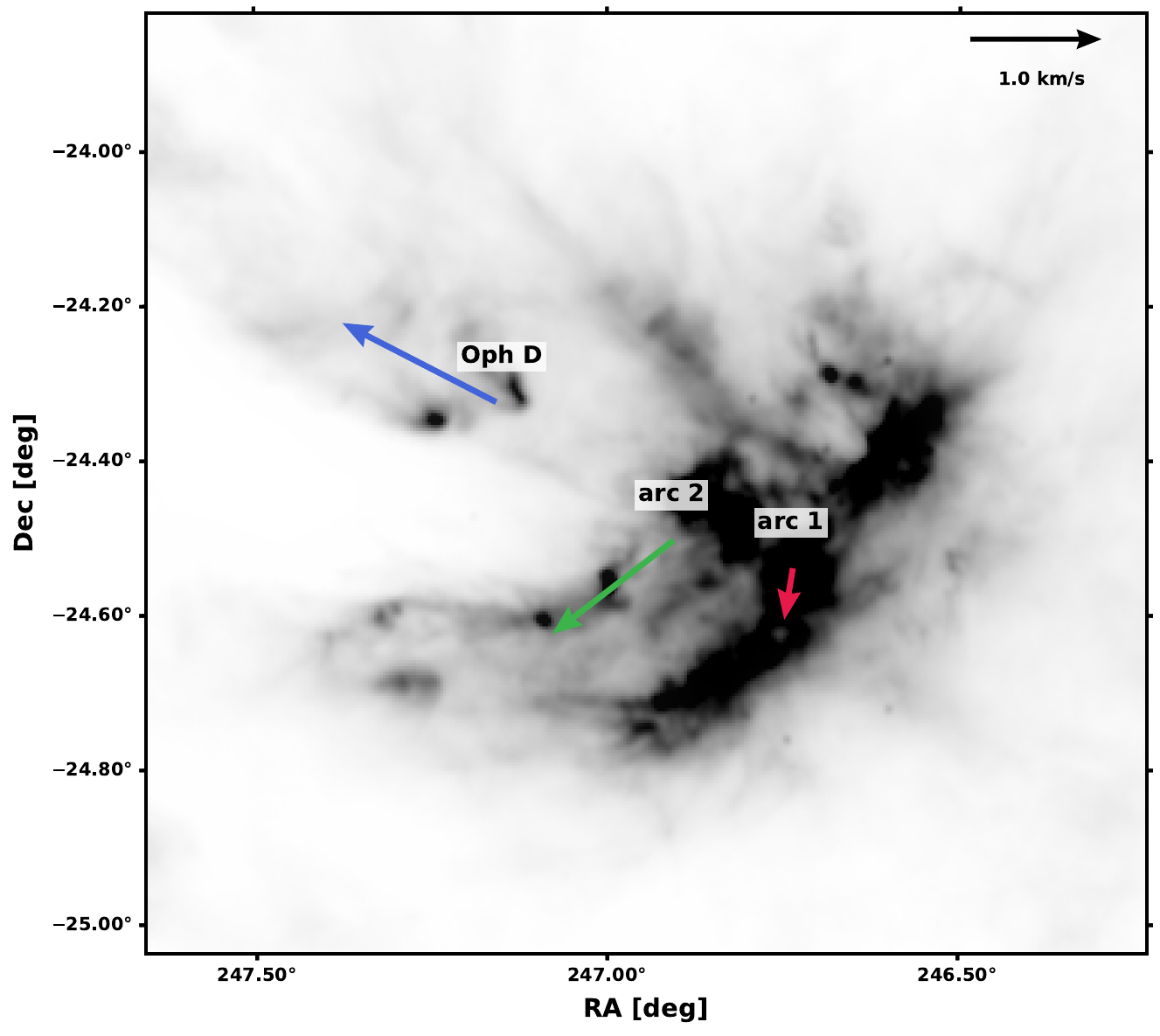}
    \caption{Velocity vectors between populations along compression fronts identified by \citet{Ladjelate_2020}.  Each vector represents the difference in perspective corrected transverse velocity with the vector rooted at the median position of population. Uncertainties on each separation velocity was estimated via bootstrapping where sources within each group was resampled 1000 times, the bulk velocity recomputed for each resample, and the standard deviation of the resulting distribution taken as the 1$\sigma$ uncertainty. The reference arrow in the top-right corner indicates a velocity scale of 1 km/s.}
    \label{fig: arcs}
\end{figure}

\subsubsection{Velocity Trends Between On \& Off Components}


To further constrain the direction of feedback influence across the cloud, we measure the velocity offset between the on and off cloud components of each subregion. We compute the bulk perspective corrected transverse velocity of both the on and off cloud and define the separation velocity as $v_{on}$ and $v_{off}$. We apply the same approach to the on- and off-cloud components of each subregion, requiring at least three sources with full kinematics on each side after clipping. Only Oph A and Oph E + F meet this requirement. Oph A shows an offset of $|\Delta\boldsymbol{v}| = 0.93^{+0.40}_{-0.28}$\,km\,s$^{-1}$, primarily in $v_{tb}$ ($2.7\sigma$; permutation $p = 0.10$), and Oph E + F an offset of $0.82^{+0.43}_{-0.20}$\,km\,s$^{-1}$ ($2.6\sigma$; $p = 0.09$). Both offsets are suggestive but not conclusive. We discuss the implications of this further in the discussion section.

\section{Discussion}
\label{dis}

\subsection{Differential Kinematics within $\rho$ Oph}

The velocity patterns across $\rho$ Oph do not exhibit a coherent global expansion pattern. Instead, the kinematic differences between subregions suggest that multiple dynamical influences may be operating across the cloud. In particular, the different motions of the two L1688 arcs, the motion of Oph D along an axis broadly parallel to its local streamer, and the more ambiguous velocity offsets in L1689 suggest that feedback from Sco OB2 has not affected all regions of $\rho$~Oph in a uniform way.

Within L1688, Arc 1 and Arc 2 show evidence for a difference in mean transverse velocity, $\Delta v_{tl} = 0.78 \pm 0.25$\,km\,s$^{-1}$ ($3.2\sigma$; permutation $p = 0.02$; Section \ref{sec:arcs}). This velocity difference is not aligned with the feedback directions indicated by the streamers, further complicating a scenario in which a single feedback source drives both arcs simultaneously. \citet{Ladjelate_2020} propose that arc-like structures in L1688 originate from multiple successive shockwave compressions propagating from the Sco OB2 association. Arc 1 and Arc 2 have similar ages, consistent with compression events occurring over a similar time interval. 


Oph D may provide a suggestive example of how large-scale feedback could interact with local cloud structure. Compared with the other arc components, Oph D shows velocities broadly aligned with the axis of the local streamer identified by \citet{Ladjelate_2020}. This geometric correspondence is consistent with a possible connection between the stellar kinematics and the local gas structure, although it does not by itself show that gas is moving along the streamer.\citet{Ladjelate_2020} find that the dense gas masses of L1709 and L1689 are lower than that of L1688 by more than a factor of two despite comparable total gas masses, meaning the material reservoir is similar but the compression status is fundamentally different. Oph D is smaller still, and its relatively low mass may make it particularly susceptible to acceleration by external feedback. The younger ages and distinct kinematics of L1709 and L1689 relative to L1688 may therefore indicate that these clouds are at a less advanced stage of externally driven compression. This could result from their greater distance from Sco OB2 and correspondingly lower incident feedback flux, or from differences in the local filamentary structure that regulate how effectively feedback couples to the gas. This picture does not, however, provide a straightforward explanation for the L1688 arcs, which may instead reflect an interaction between the two arcs (Section\ref{sub: col}). The contrasting kinematics of L1688, Oph D, and L1689 suggest that the response of $\rho$~Oph to external feedback was neither spatially uniform nor ordered along the direction of propagation proposed by \citet{Ladjelate_2020}. We therefore find that the structure of $\rho$ Oph is not consistent with the influence of feedback alone.

While feedback from Sco OB2 may have contributed to the dynamical evolution of the cloud, the stellar kinematics are more consistent with a combination of feedback, local cloud structure, and distinct star-forming processes than with a single coherent sequence of feedback-driven compression events.

\subsection{Connecting the Arcs \& Subregions to the Upper-Sco Chain}
\label{sub: col}
\citet{Posch_2025} decompose Upper Sco into two chains of sequentially triggered star formation: Upper-Sco I ($\phi$ Lup, $\eta$ Lup, Sco-Body, $\rho$ Sco and Antares) and Upper-Sco II ($\delta$ Sco, $\sigma$ Sco, $\beta$ Sco and $\nu$ Sco). We adopt this framework as a hypothesis and test whether the $\rho$~Oph substructures are compatible with it. $\rho$ Oph sits at the young end Upper-Sco I, with a notably larger membership than the preceding clusters in either chain, which \citet{Posch_2025} propose may indicate the onset of a new generation of triggered star formation. 

\begin{figure*}[t]
    \centering
    \includegraphics[width=\textwidth]{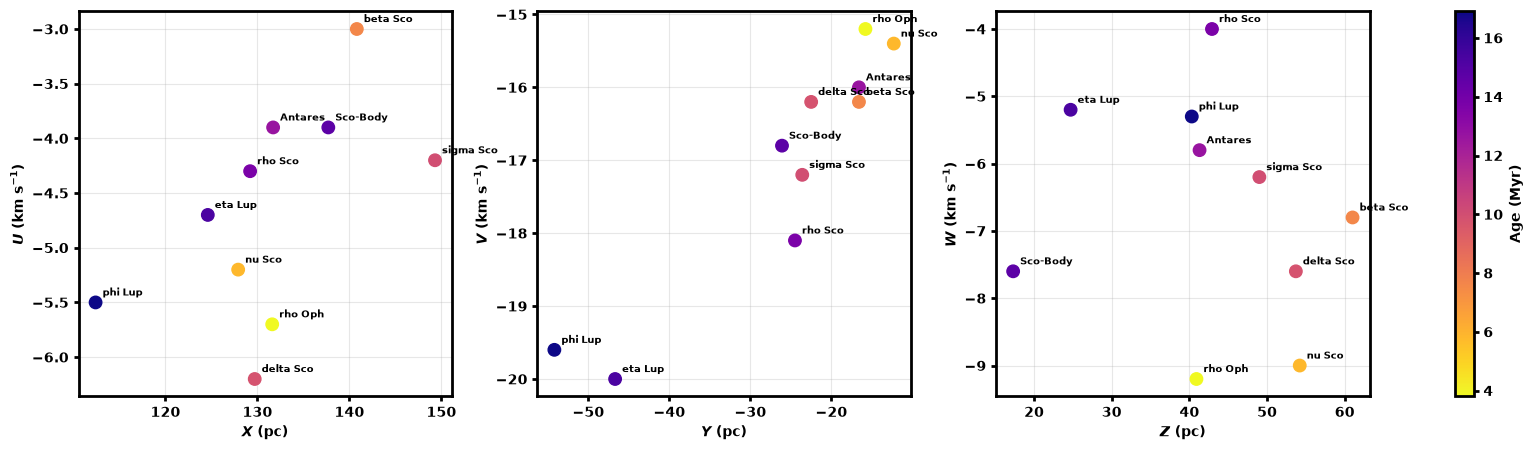}
    \caption{Heliocentric UVW velocity components of the Upper Sco cluster populations from \citet{Posch_2025} plotted against their respective Galactic Cartesian position axes (X, Y, Z), coloured by cluster age. All positions and velocities are taken directly from \citet{Posch_2025}.}
    \label{fig:posch}
\end{figure*}

Figure \ref{fig:posch} shows UVW velocity components of each cluster plotted against their respective Galactic Cartesian position axes, coloured by age. If Upper-Sco were a single coherent expanding population diverging from a common origin, we would expect a positive linear gradient of velocity with position along a given axis. The V-Y panel shows the clearest such trend, while the U-X panel is considerably noisier with notable outliers, and the W-Z panel shows a negative trend indicating convergence. Separating the populations into Upper-Sco I and Upper-Sco II chains produces more positive trends in the Z direction, suggesting the apparent convergence may partly results from mixing distinct kinematic substructures and inconsistent with a simple expansion.

To test for interactions between L1688 and Upper Sco, we traced back the orbits of Arc 1, Arc 2 and Cluster 0, together with each \citet{Posch_2025} population, for 20\,Myr using \texttt{galpy} with \texttt{MWPotential2014} \citep{Bovy_2015}. Group positions and velocities are medians over the members used in the velocity analysis. We ran 1000 Monte Carlo realizations, perturbing each member's distance, proper motion and radial velocity within its uncertainties, bootstrap resampling the members, and perturbing the \citet{Posch_2025} medians within their uncertainties. We quote the median distance ($d_{\rm min}$) and time ($t_{\rm min}$) of closest approach, with 16th--84th percentile ranges. Both arcs trace back through Upper Sco~II (Table \ref{tab:traceback_arcs}). Their closest approach is to $\delta$ Sco, which also contains our Cluster 0: Arc 1 passes within $4.9^{+1.7}_{-1.8}$\,pc $5.8^{+0.5}_{-0.4}$\,Myr ago, and Arc 2 within $5.4^{+3.6}_{-2.6}$\,pc $6.6^{+1.5}_{-1.5}$\,Myr ago. Neither arc is associated with $\delta$ Sco alone, however. Both also pass $\beta$ Sco at 8--9\,pc and $\sigma$ Sco at 12--15\,pc over the same 4--6\,Myr interval (Table \ref{tab:traceback_arcs}). 

The two arcs' approaches agree within their uncertainties, which are dominated by the radial direction. Moreover, both arcs approached several Upper Sco~II populations rather than a single source, suggesting that the cloud was influenced by successive generations of feedback rather than one compression event. These encounters occurred ${\sim}5$\,Myr before the arcs began forming stars, too early to have triggered the present star formation directly, but consistent with Upper Sco II feedback having shaped the cloud beforehand.  The arcs nevertheless have distinct velocities, differing by ${\approx}0.8$\,km\,s$^{-1}$. In Figure \ref{fig: arcs}, where velocities are measured relative to Group~8, Arc~2 has the larger velocity, but relative to the Upper Sco~II populations the order reverses: Arc~1 is receding from $\delta$ Sco at ${\sim}2.3$\,km\,s$^{-1}$, compared with ${\sim}1.9$\,km\,s$^{-1}$ for Arc~2, away from a shell that is prominent in gas and dust maps north of L1688. Since both arcs passed close to all of the Upper Sco~II populations, one explanation is that they contain residual gas from different generations of Upper Sco~II that received different feedback kicks, with the younger component receiving the larger kick and catching up with the other. Alternatively, the arcs may simply have inherited slightly different velocities; the offset is less well aligned in $Y$ and $Z$.

\begin{deluxetable*}{lcccccc}
\tabletypesize{\scriptsize}
\tablecaption{Closest approaches between the L1688 arcs and the Upper Sco populations of \citet{Posch_2025}.\label{tab:traceback_arcs}}
\tablehead{
\colhead{} & \multicolumn{2}{c}{Upper Sco I} & \multicolumn{4}{c}{Upper Sco II} \\
\cline{2-3} \cline{4-7}
\colhead{Group} & \colhead{$\rho$ Sco} & \colhead{Antares} & \colhead{$\sigma$ Sco} & \colhead{$\delta$ Sco} & \colhead{$\beta$ Sco} & \colhead{$\nu$ Sco}
}
\startdata
Arc 1 & $5.9^{+0.6}_{-0.6}$\,($1.1^{+0.1}_{-0.1}$) & $1.6^{+0.6}_{-0.6}$\,($0.7^{+0.1}_{-0.1}$) & $15.4^{+1.3}_{-1.4}$\,($4.1^{+0.3}_{-0.2}$) & $4.9^{+1.8}_{-1.8}$\,($5.8^{+0.5}_{-0.4}$) & $8.4^{+1.4}_{-1.2}$\,($5.5^{+0.4}_{-0.4}$) & $11.2^{+2.1}_{-2.4}$\,($10.5^{+2.9}_{-2.6}$) \\
Arc 2 & $7.9^{+2.2}_{-1.4}$\,($0.9^{+0.3}_{-0.4}$) & $1.7^{+2.4}_{-0.7}$\,($0.5^{+0.4}_{-0.5}$) & $11.7^{+2.4}_{-2.4}$\,($4.4^{+0.8}_{-0.9}$) & $5.4^{+3.6}_{-2.6}$\,($6.6^{+1.5}_{-1.5}$) & $9.4^{+2.3}_{-2.2}$\,($5.1^{+0.8}_{-1.0}$) & $13.6^{+2.1}_{-3.6}$\,($6.6^{+4.7}_{-6.0}$) \\
\enddata
\tablecomments{The table shows the median closest-approach distance $d_{\rm min}$ (pc) and, in parentheses, the time since closest approach $t_{\rm min}$ (Myr ago) over 1000 Monte Carlo realizations, with errors from the 16th and 84th percentiles. $\phi$ Lup, $\eta$ Lup and Sco-Body are omitted, as neither arc approaches them within 24\,pc.}
\end{deluxetable*}

This makes specific predictions that would falsify it if absent. If Arc 1 and Arc 2 are interacting, the molecular gas should show (1) two coherent velocity components associated with the two arcs; (2) enhanced column densities where they meet; and (3) an interaction timescale. Position-velocity cuts through $^{13}$CO data, or higher-resolution observations of dense-gas and shock tracers, could test these directly. 

\subsubsection{Interpreting relative ages in a dynamically complex region}
\label{sec:relative_ages}

An important consideration in interpreting the age gradients in $\rho$ Oph is that the different age diagnostics used here do not necessarily share the same absolute calibration. In particular, pre-main-sequence isochronal ages depend on the adopted stellar evolutionary models, and empirical tests using eclipsing binaries have demonstrated significant model-dependent discrepancies, especially for low-mass stars \citep{Stassun2014}. These differences are directly relevant to Upper Sco: eclipsing-binary masses and radii yield ages of approximately 5--7 Myr with standard evolutionary models, but 9--10 Myr when magnetic effects are included \citep{David2019}. The class-based ages of the embedded $\rho$ Oph populations carry a different set of assumptions, notably the adopted durations of the protostellar and disk-bearing phases and the underlying star-formation history. Thus, the absolute age contrast between embedded and optically accessible populations should not be interpreted as a uniformly calibrated clock.

Nevertheless, the relative spatial trends provide useful constraints. In particular, the tendency for off-cloud populations to be older than their on-cloud counterparts, together with the similar characteristic ages but distinct transverse motions of the two L1688 arcs, is consistent with a region assembled through more than one episode or pathway of star formation. The kinematic distinction need not imply a correspondingly resolvable age difference: distinct gas structures may form stars over overlapping intervals, and subsequently preserve different bulk motions. Conversely, a monotonic age gradient alone does not uniquely establish propagation of a triggering front. The strongest evidence for a causal feedback sequence would require consistency among the relative ages, the three-dimensional motions of the stars and gas, and the timing and geometry of candidate feedback sources. In this sense, the combined demographic and kinematic analysis presented here is more informative than either diagnostic in isolation, while appropriately leaving open whether feedback, cloud interactions, or both produced the observed substructure.
\section{Conclusions}
\label{sec:conc}

In this work, we construct a comprehensive catalog of protostars, dense cores, and pre-main-sequence stars in $\rho$ Oph, and characterize the ages and kinematics of its subregions. We combine clustering, isochronal ages, on-sky velocities and dynamical traceback with the sequentially formed Upper Sco populations identified by \citet{Posch_2025} to investigate the origin of $\rho$ Oph's large stellar population. In particular, we test whether its present-day structure and kinematics are consistent with a scenario in which successive generations of massive stars in Upper Sco drove feedback into the surrounding clouds, producing sequential episodes of star formation that culminates in the young and populous $\rho$ Oph complex. Our key findings can be summarized as follows:

\begin{enumerate}

\item We find a broad west-to-east age gradient across the complex, with the oldest population (Cluster 0, $\sim$7.94 Myr) in the west, progressively younger ages toward the L1688 core ($\sim$1 Myr), and the youngest populations in L1709 and L1689. Most subregions host distinct off-cloud populations that have decoupled from their natal gas, typically $\sim$1 Myr older than their on-cloud counterparts.


\item Both L1688 arcs are diverging from the Upper Sco~II populations they passed several Myr before forming stars. The arcs have distinct velocities, offset mostly along the filaments in a way consistent with a glancing interaction.This offset is difficult to explain with a single feedback-driven compression, but is more consistent with the arcs containing gas from different generations of Upper Sco~II that received different feedback kicks.

\item The spatial and kinematic properties of the subregions show that the connection to the proposed Upper-Sco feedback sequence is not uniform across $\rho$ Oph. L1709 and L1689 show convergent velocities relative to older subgroups, distinct from the kinematics of L1688, while Oph D is the clearest case where the observed population is consistent with proposed feedback direction. Together, these differences indicate that subregions may have experienced distinct dynamical influences rather than a single coherent evolutionary sequence. In L1689, the observed convergent motion may represent a related stage of compression history and mediate continued star formation.

\end{enumerate}

Our results reveal $\rho$ Oph as both dynamically and temporally heterogeneous, whose structure reflects the cumulative effects of potentially multiple instances of feedback and cloud-cloud interactions. Our confidence in this picture is limited by the lack of reliable proper motions for the deeply embedded population, which dominates the region's stellar census but remains kinematically under-characterized. The Vera Rubin Observatory and eventually, Gaia NIR will provide an important complement by extending kinematic observations into the heavily obscured regions of $\rho$ Oph and other nearby star-forming regions.

\section{Acknowledgements}

SEA was supported by the Summer Undergraduate Research Program (SURP) at the University of Toronto under the supervision of RMPK, who is funded as a Dunlap Fellow at the Dunlap Institute for Astronomy and Astrophysics. The Dunlap Institute is funded through an endowment established by the David Dunlap family and the University of Toronto. Partial support for the contributions of TR and LP came from National Science Foundation awards AST-2206443 (to TR) and AST-2206703/AST-2630422 (to LP). 

This project uses unreleased data from the Sloan Digital Sky Survey V (SDSS-V). Funding for SDSS-V has been provided by the Alfred P. Sloan Foundation, the Heising-Simons Foundation, the National Science Foundation, and the Participating Institutions. SDSS acknowledges support and resources from the Center for High-Performance Computing at the University of Utah. SDSS telescopes are located at Apache Point Observatory, funded by the Astrophysical Research Consortium and operated by New Mexico State University, and at Las Campanas Observatory, operated by the Carnegie Institution for Science. The SDSS web site is \url{www.sdss.org}.

SDSS is managed by the Astrophysical Research Consortium for the Participating Institutions of the SDSS Collaboration, including the Carnegie Institution for Science, Chilean National Time Allocation Committee (CNTAC) ratified researchers, Caltech, the Gotham Participation Group, Harvard University, Heidelberg University, The Flatiron Institute, The Johns Hopkins University, L'Ecole polytechnique f\'{e}d\'{e}rale de Lausanne (EPFL), Leibniz-Institut f\"{u}r Astrophysik Potsdam (AIP), Max-Planck-Institut f\"{u}r Astronomie (MPIA Heidelberg), Max-Planck-Institut f\"{u}r Extraterrestrische Physik (MPE), Nanjing University, National Astronomical Observatories of China (NAOC), New Mexico State University, The Ohio State University, Pennsylvania State University, Smithsonian Astrophysical Observatory, Space Telescope Science Institute (STScI), the Stellar Astrophysics Participation Group, Universidad Nacional Aut\'{o}noma de M\'{e}xico, University of Arizona, University of Colorado Boulder, University of Illinois at Urbana-Champaign, University of Toronto, University of Utah, University of Virginia, Yale University, and Yunnan University.

\begin{contribution}

This paper was led by, written by, and submitted by SEA. RMPK supervised the project as part of the Summer Undergraduate Research Program (SURP) and later an instructor for the University of Toronto's AST 430 guided research course, providing guidance on the overarching methodologies and composition of the paper, and recommending revisions throughout. ARL, SSS, JGFT, and RRL joined through the SDSS collaboration and contributed detailed reviews that significantly improved the final version of this paper. KGS also joined through SDSS, and contributed subsection 5.2.1. LP and TAR contributed knowledge of the distribution of Herbig-Haro objects in the region, which are being studied in more depth through a sister paper, and also assisted with final editing of the manuscript. MK is an architect in the SDSS collaboration, and is included due to her contributions to the data that this publication relies on. 

\end{contribution}

\section{Code Acknowledgments}
Galpy \citep{Bovy_2015}, Astropy \citep{astropy_2022}, NumPy \citep{numpy}, Pandas \citep{pandas}, SciPy \citep{scipy}, Matplotlib \citep{Hunter_2007}, Pyutils (cite), HDBSCAN \citep{HDBSCAN}

\bibliography{bibliography}{}
\bibliographystyle{aasjournalv7}

\end{document}